\documentclass[aps,prd,twocolumn,superscriptaddress,preprintnumbers,floatfix,nofootinbib,longbibliography]{revtex4-2}
\usepackage{hyperref}

\usepackage{blindtext}

\usepackage{mathtools,mathalfa,amssymb,amsmath,amsfonts,amsthm,bm}
\usepackage[T1]{fontenc}
\usepackage{tipa}
\usepackage{wasysym}
\usepackage{appendix}
\usepackage[normalem]{ulem}
\usepackage{calrsfs}
\usepackage{tikz}
\usetikzlibrary {decorations.pathmorphing}
\usepackage{graphicx}

\usepackage[font=small]{caption}
\usepackage{subcaption}

\usepackage[mathscr]{euscript}

\renewcommand{\d}{\mathrm{d}}
\newcommand{\re}{\mathrm{Re}}

\newcounter{definition}
\definecolor{wuppergreen}{RGB}{137, 186, 23}
\newcommand{\pullback}[1]{\hbox{\lower0.5ex\hbox{${}_{\leftarrow}$}}\kern-1.9ex{#1}}
\newcommand{\pullbacklong}[1]{\hbox{\lower0.85ex\hbox{${}_{\longleftarrow}$}}\kern-3.0ex{#1}}
\newcommand{\pullbackllong}[1]{\hbox{\lower0.85ex\hbox{${}_{\longleftarrow\!\!-\!\!-\!\!-\!\!-}$}}\kern-6.4ex{#1}}

\begin{document}

\title{Universality in the Transition from Inspiral to Plunge {for Extreme-Mass-Ratio-Inspirals}:
High-Accuracy Analytic Solutions and Catastrophe Theory}
\author{Ariadna Ribes Metidieri}
\email{ariadna.metidieri@nbi.ku.dk}
\affiliation{Center of Gravity, Niels Bohr Institute, Blegdamsvej 17, 2100 Copenhagen, Denmark}
\affiliation{Max-Planck-Institut f{\"u}r Gravitationsphysik (Albert
  Einstein Institute), Callinstra{\ss}e 38, 30167 Hannover, Germany}
\affiliation{Institute for Mathematics, Astrophysics and Particle Physics, Radboud University, Heyendaalseweg 135, 6525 AJ Nijmegen, The Netherlands}

\author{B\'eatrice Bonga}
\email{bbonga@science.ru.nl}
\affiliation{Institute for Mathematics, Astrophysics and Particle Physics, Radboud University, Heyendaalseweg 135, 6525 AJ Nijmegen, The Netherlands}

\author{Badri Krishnan} \email{badri.krishnan@ru.nl}
\affiliation{Institute for Mathematics, Astrophysics and Particle
  Physics, Radboud University, Heyendaalseweg 135, 6525 AJ Nijmegen,
  The Netherlands}
\affiliation{Max-Planck-Institut f{\"u}r Gravitationsphysik (Albert
  Einstein Institute), Callinstra{\ss}e 38, 30167 Hannover, Germany}
\affiliation{Leibniz Universit{\"a}t Hannover, 30167 Hannover,
  Germany}
  \author{Jos\'e Luis Jaramillo}
  \email{Jose-Luis.Jaramillo-Martin@ube.fr}
  \affiliation{Institut de Math\'ematiques de Bourgogne UMR 5584,
    Universit\'e Bourgogne Europe, CNRS, F-21000 Dijon, France}

\begin{abstract}
We revisit the transition from inspiral to plunge for extreme mass-ratio inspirals on quasi-circular, inclined orbits in Kerr spacetime from the perspective of catastrophe theory. Our goal is to uncover the mathematical structures underlying the universality of the transition dynamics, which remains governed by the same Painlevé I differential equation as for equatorial inspirals despite the additional complexity.
We first analyze the solution of the Painlevé I equation selected by the physical boundary conditions of slowly evolving quasi-circular inspiral at early times. We argue that these conditions uniquely select the 
{ so-called Boutroux’s} \emph{tritronquée solution} of Painlevé I. 
We then compare existing high-accuracy analytic approximations of the tritronquée solution with direct numerical integrations of the Painlevé I equation, finding comparable accuracy and improved stability under differentiation and integration for the analytic solution.
In the second part of this work, we show that the equilibrium structure of the Kerr radial effective potential admits a natural interpretation in terms of catastrophe theory. Equatorial orbits are associated with the fold catastrophe, while inclined orbits are described by the cusp catastrophe. In both cases, the transition to plunge corresponds to slow evolution across fold lines of the catastrophe manifold, providing a geometric explanation for the universal appearance of the Painlevé I equation in the transition dynamics.

\end{abstract}

\maketitle
\section{Introduction}
The problem of binary black hole mergers in general relativity has
deservedly attracted significant attention in gravitational wave (GW)
astronomy. A gravitationaly bound system of two compact objects loses
energy as it emits gravitational radiation.  This emitted radiation
carries information about the dynamics and the ultimate fate of the
binary system as the two black holes eventually merge and form a
single remnant black hole.  This entire process of the inspiral and
merger of two black holes have been observed by current generation
gravitational wave detectors; see e.g. {\cite{LIGOScientific:2026wfs}}
for the most recent observational results. These merger events carry a
wealth of information regarding gravity in the dynamical strong-field
regime, and some of these events already allow stringent tests of
general relativity.  While the inspiral and post-merger regimes of a
compact binary merger are amenable to analytical treatments, these
methods are not directly applicable to the merger regime, which
currently requires numerical calculations for an accurate description.
Thus, it is the merger regime which might be expected to harbor
new, perhaps as yet undiscovered, phenomena pertaining to the
dynamical non-perturbative aspects of general relativity.

Future generations of GW observatories on Earth
\cite{Evans:2021gyd,Reitze:2019iox,ET:2019dnz,Sathyaprakash:2012jk}
and in space \cite{LISA:2024hlh,TianQin:2015yph,Hu:2017mde} will allow
us to measure the emitted GW signal with great
precision.  One of the important GW sources for space-based
observatories are extreme mass ratio inspirals (EMRIs)
\cite{Babak:2017tow}.  While the current generation of GW detectors
are most sensitive to comparable stellar-mass binaries, space-based detectors such as LISA are sensitive at lower frequencies
corresponding to more massive systems. These include the highly
asymmetric EMRIs consisting of a stellar mass compact object captured
by a supermassive black hole.

A typical EMRI evolution can be divided into three regimes: the adiabatic inspiral, the transition-to-plunge and the plunge. Roughly speaking, the corresponding regimes for comparable-mass systems would be inspiral, merger and ringdown. The transition-to-plunge phase carries little signal-to-noise (SNR) in LISA and other space-based gravitational-wave detectors \cite{Ori:2000zn} and has therefore often been treated only approximately or ignored in waveform pipelines. Nonetheless, the transition-to-plunge becomes significantly more relevant for intermediate-mass-ratio inspirals (IMRIs) \cite{Huerta:2014irs,Kuchler:2024esj}, (partially) explaining the recent interest in modeling this regime more accurately \cite{Apte:2019,Lim:2019xrb,Compère_2020-Painleve,Compere:2021zfj,Compere:2021iwh}. This regime is also crucial for connecting the long adiabatic inspiral to the final ringdown with controlled phase and amplitude accuracy. A robust description of the transition is therefore essential for performing precision tests of black holes and general relativity that rely on coherent matching of inspiral, plunge and ringdown, or when extracting strong-field parameters from IMRI signals where the transition can contribute non-negligibly to the observable waveform.

In this work we re-examine the transition from inspiral to plunge for quasi-circular, inclined EMRIs.
Just as for comparable mass systems, while there exist
analytical methods which accurately decribe the inspiral (and ringdown) regimes, the merger remains less well understood.  There are
two principal difficulties:
\begin{enumerate}
\item The merger may have non-perturbative aspects.  This is especially
  true for comparable mass systems wherein the spacetime in the
  vicinity of the merger cannot be treated as a linear perturbation of
  a Kerr black hole.  
\item In the inspiral regime there is a natural separation of time
  scales allowing us to treat the problem as an adiabatic evolution;
  the radiation reaction which is responsible for the shrinking of the
  orbit, operates on a longer timescale than the orbital motion.
  However, in the merger regime, this separation of timescales is no
  longer possible and other methods are required.
\end{enumerate}
EMRIs are simpler than comparable mass systems in the sense that, even at the merger, the spacetime can always be considered to be a
perturbation of a background Kerr spacetime corresponding to the
central supermassive black hole.  EMRIs therefore provide us with a simplification (while
still remaining very interesting for GW astronomy). The second difficulty, namely dealing with
non-adiabatic dynamics at the merger, is also more straightforward for EMRIs as the transition-to-plunge still has a separation of scales, albeit a different one from the adiabatic inspiral. 

Building on earlier derivations for equatorial motion \cite{Ori:2000zn,Buonanno:2000ef} and recent extensions to inclined Kerr orbits {\cite{Apte:2019}}, we stress that the transition to plunge equation reduces to the Painlevé I equation, as was already pointed out in \cite{Compère_2020-Painleve} for quasi-circular equatorial orbits, and in \cite{Apte:2019} for inclined orbits. { Solutions to Painlevé equations are called ``Painlevé transcendents'' and depend on the imposed boundary and/or asymptotic conditions. In the case of equatorial motion, it was  { identified} in Ref. \cite{Jaramillo:2022oqn} that under
the physical boundary condition that the inspiral is slowly evolving and quasi-circular at early times,  the real Boutroux's tritronquée Painlevé I transcendent is uniquely selected. In the present work, we { prove this result in a systematic manner}, and also extends { it} to inclined orbits.  Moreover, } identifying the physical solution with the tritronquée one not only fixes the solution's analytic structure (in particular the location of the first real pole that signals the breakdown of the transition approximation), it also permits the use of a recent high-accuracy, uniformly valid analytic approximation to the Painlev\'e I equation \cite{ADALITANVEER20163843}. This high-accurate analytic solution, applied for the first time to the gravitational wave problem, also comes with rigorous (and uniform) error bounds, and we show that its performance is comparable to that of a numerical implementation, while retaining the advantage of a closed-form analytical expression. Therefore, for waveform modeling, this solution should be used.

Independently, we interpret the transition-to-plunge regime in the language of catastrophe theory \cite{Berry_1981,poston1996catastrophe}. Catastrophe theory is a powerful tool that, { in particular,} classifies the structurally stable ways in which critical points of a smooth potential are created or annihilated as control parameters evolve slowly. In our setting, the relevant potential is the effective potential describing the radial motion, the control parameters are combinations of the ``constants'' of motion describing the trajectory, viz. energy, angular momentum and the Carter constant, and the critical points mark the transition-to-plunge. In GW science catastrophe theory has already been { used, for example,  in the exploration of universality and simplicity features of the binary black hole merger waveform,} to model  features associated with fold/cusp caustics in electromagnetic and gravitational-wave lensing, to improve waveform modeling in certain regimes of Fourier space  and to define orbital eccentricity \cite{Jaramillo:2022mkh, Bonga:2024orc,Serra:2025kbw,Ezquiaga:2025gkd,Loutrel:2023rsl,Boschini:2024scu}. { In a related gravitational setting, catastrophe or singularity theory  has also been recently used to study { non-differentiable stable structures in} black hole horizons \cite{Gadioux:2023pmw,Gadioux:2024tlm,menesesrojas:tel-05553646}.} 
%For us, the radial effective potential defines the `catastrophe or solution manifold' {\jl in such a way that the generic singularities of the projection from such `catastrophe manifold' to the `control manifold' --- known as catastrophes,  ---} are folds (equatorial motion) and cusps (inclined motion),
For us, the radial effective potential serves as a starting point to apply the framework of catastrophe theory. As we shall see in Sec.~\ref{subsec:catastrophe-theory}, the transition-to-plunge is governed by the fold (equatorial orbits) and cusp (inclined orbits) catastrophes.   
%For us, the radial effective potential defines the catastrophe manifold {\jjl and its projection to the so-called `state space', namely the `catastrophe map'}, whose generic singularities --- known as catastrophes --- are folds (equatorial motion) and cusps (inclined motion);
%for more details see section \ref{subsec:catastrophe-theory}. 
%Despite the more general catastrophe in the case of inclined orbits, 
The transition-to-plunge dynamics always selects a slow radiation-reaction drift crossing a fold.
%line (except for a special case in which the black hole is extremal and the orbit has a particular inclination angle).
This immediately explains the observed universality for equatorial and inclined orbits.%: near the fold, the dynamics is governed by the fold normal form and therefore reduces locally to the same Painlev\'e~I equation.

The present work is related to the program to understand the simplicity and universality features of the binary black hole merger waveform sketched in \cite{Jaramillo:2022mkh,Jaramillo:2022oqn,Jaramillo:2023day,JarAlSBes26}.  However, in  contrast with the qualitative and heuristic tone of the discussion in those references, the present work aims at addressing the concrete problem of plunge dynamics.
%\footnote{\jjl In this sense, the Painlevé I part of this work could be related to the third layer (focused on Painlevé transcendents and EMRIs) of the hierarchy
%of asymptotic models  proposed in the respective
%tables of \cite{Jaramillo:2023day,JarAlSBes26}, aiming at studying binary black
%hole dynamics.}, in a thorough and detailed treatment.
%}

The remainder of the paper is organized as follows. In Sec.~\ref{sec:background} we review Kerr geodesics, define quasi-circular inclined orbits and the ISSO, and re-derive the leading-order transition equation. In Sec.~\ref{sec:Ori-Thorne-Kesden} we identify the physical solution with the tritronquée Painlevé I transcendent and present the explicit high-accuracy approximation together with its error bounds. There we also compare the analytic approximation with numerical integrations and discuss implications for waveform modeling. Sec.~\ref{sec:universality} develops the catastrophe-theoretic interpretation and analyzes structural stability of the Painlev\'e I equation. We conclude in Sec.~\ref{sec:discussion} with a summary and outlook.

\section{Background}\label{sec:background}

In an EMRI, the small but finite mass of the compact object ($\mu$) gives rise to radiation reaction, leading to the gradual loss of energy and angular momentum through gravitational-wave emission. During the inspiral, the motion can be described as an adiabatic sequence of geodesics with slowly evolving orbital parameters. As the system approaches the {last stable orbit}, the separation between the orbital and radiation-reaction timescales breaks down. The adiabatic approximation then fails, and a distinct perturbative treatment is required to model the transition from inspiral to plunge. The plunge phase itself is well approximated by geodesic motion in the background spacetime.

In this work, we provide an analytic solution to the trajectory of a small non-spining and non-precessing compact object from the transition regime up to the final plunge, to leading order in the mass-ratio. Moreover, we provide a description of the transition using the language of catastrophe theory. We will consider a Kerr spacetime with an orbit of arbitrary inclination. However, we restrict our attention to quasi-circular orbits, i.e., orbits with vanishing eccentricity~\footnote{Although some of the methods and ideas developed in this work may prove useful for eccentric inspirals, they are not directly applicable in that case.}. 
Hence, instead of the Last Stable Orbit, the transition will take place around the {innermost stable circular orbit} (ISCO) for equatorial motion, or around the {innermost stable spherical orbit} (ISSO), its generalization for inclined orbits. Throughout this work, we use geometrized units ($G=c=1$). 

In the remainder of this section, we review the status of previous work on this topic, which is also used to set the notation. In Sec.~\ref{subsec:Geodesics-kerr}, we review the timelike geodesic equation in a Kerr background, and the location of the ISSO for quasi-circular, inclined orbits. In Sec.~\ref{subseec:transition-to-plunge} we re-derive the differential equation governing the transition-to-plunge for inclined orbits in a Kerr background. 

\subsection{Geodesics in the Kerr background }
\label{subsec:Geodesics-kerr}

The geodesic motion in Kerr spacetime (a Petrov type D solution) is completely integrable due to the existence of a second-rank Killing tensor, which gives rise to the Carter constant~\cite{Carter:1968rr}. As a consequence, there exist four constants of motion that allow the geodesic equations in a Kerr background with mass $M$ and spin $a$ to be written as a system of four first-order coupled nonlinear differential equations.

Introducing Mino time $\lambda$, defined by $\mathrm{d}\lambda = \mathrm{d}\tau/\Sigma$ with $\Sigma = r^2 + a^2 \cos^2\theta$~\cite{Mino:2003yg}, decouples the radial and polar equations. The geodesic equations in Boyer-Lindquist coordinates then take the form
\begin{equation}
\label{eq:geodesic-Kerr-Mino}
    \begin{aligned}
\left(\frac{d r}{d \lambda}\right)^2 & =  {\left[E\left(r^2+a^2\right)-a L_z\right]^2 } \\
& -\Delta\left[r^2+\left(L_z-a E\right)^2+Q\right] \\
& \equiv  R(r)\,, \\
\left(\frac{d \theta}{d \lambda}\right)^2 & =  Q-  L_z^2 \cot ^2 \theta-a^2 \cos ^2 \theta\left[1-E^2\right] \\
& \equiv  \Theta(\theta)\,,\\
\left(\frac{d \phi}{d \lambda}\right) & =\csc ^2 \theta L_z+\frac{2 M r a E}{\Delta}-\frac{a^2 L_z}{\Delta} \\
& \equiv \Phi(r, \theta)\,,
\\
\left(\frac{d t}{d \lambda}\right) &   = E\left[\frac{\left(r^2+a^2\right)^2}{\Delta}-a^2 \sin ^2 \theta\right]-\frac{2 M r a L_z}{\Delta} \\
& \equiv T(r, \theta)\,,
\end{aligned}
\end{equation} where $\Delta = r^2-2M r+a^2$, $E$ is the orbital energy (per unit $\mu$), $L_z$  the $z$-component of the orbital angular momentum (per unit $\mu$), and $Q$ the Carter constant (per unit $\mu^2$). 
To avoid notational clutter, we will also denote $L_z$ simply as $L$ throughout the rest of the paper, as nowhere will we use the norm of the total angular momentum.

Since we restrict our attention to the study of the transition up to the final plunge, we consider bound orbits, which have $0<E<1$. Furthermore, for bounded orbits the radial function $R(r)$ can be re-written as 
\begin{equation}\label{eq:R(r)-bounded}
    R(r) = (1-E^2) (r_1-r)(r_2-r)(r_3-r)(r-r_4)
\end{equation} where $r_i\,,\quad i=1...4$ are the real roots of the fourth degree polynomial $R(r)$ such that $r_1\geq r_2\geq r_3\geq r_4$. Bound motion occurs for $r_p=r_2\leq r\leq r_1=r_a$, with $r_p$ (pericenter) and $r_a$ (apocenter) the minimum and maximum radii of the orbit. Orbits with $L>0$ ($L<0$) are prograde (retrograde), i.e., they rotate in the same (opposite) sense as the black hole spin.

Rewriting the polar equation in Eq.~\eqref{eq:geodesic-Kerr-Mino} using $\zeta=\cos\theta$ gives
\begin{equation}
    \left(\frac{\d \zeta}{\d\lambda}\right)^2 = (1-\zeta^2) Q - \zeta^2\left[L^2+(1-E^2) (1-\zeta^2)a^2\right]\,,
\end{equation} 
which shows that the Carter constant must be positive $Q>0$ for bound non-equatorial motion, while equatorial orbits (with $\zeta=0$) correspond to $Q=0$. 

    In the form of Eq.~\eqref{eq:geodesic-Kerr-Mino}, the geodesic equations can be solved hierarchically by first integrating the radial and polar equations. The temporal and azimuthal components may then be obtained by separation, writing $t(r(\lambda),\theta(\lambda),\lambda)=t_r(r(\lambda))+t_\theta(\theta(\lambda))$ and $\phi(r(\lambda),\theta(\lambda),\lambda)=\phi_r(r(\lambda))+\phi_\theta(\theta(\lambda))$ (see, e.g., Ref.~\cite{Dyson:2023fws}). In this section we therefore focus on the radial and polar motion.

In an EMRI, radiation reaction induces a slow time dependence in the constants of motion $(E,L,Q)$. Hence, the inspiral and transition are not accurately described by Eq.~\eqref{eq:geodesic-Kerr-Mino}, which applies to geodesic motion. However,  the motion is well approximated by a geodesic with fixed constants of motion during the plunge.

In this work, we restrict ourselves to quasi-circular orbits following the exposition in Refs.~\cite{Ori:2000zn,Apte:2019}. Circular orbits are defined by the conditions
\begin{equation}
\label{eq:circularity-condition}
R(r_o)=0\,, \qquad R'(r_o)=0\,,
\end{equation}
where the prime denotes partial differentiation with respect to $r$ and $r_o$ the location of a specific circular orbit. These conditions ensure that the radius remains constant at $r=r_o$.

Imposing the circularity condition yields a two-parameter family of solutions, which may be parametrized by the orbital radius $r_o$ and an inclination angle $I$. 
For $I=0$ the orbit is equatorial ($\theta=\pi/2$) and remains confined to the equatorial plane. For non-equatorial orbits, the polar motion is bounded between $\theta\in[\theta_{\text{min}},\theta_{\text{max}}]$~\cite{Apte:2019}, where 
\begin{equation}
    \theta_{\text{min}} = \text{sgn}(L) \times\left(\frac{\pi}{2}-I\right)\,,\quad \theta_{\text{max}} = \pi-\theta_{\text{min}}\,.
\end{equation}

Since circular orbits are fully specified by $(r_o,I)$, the constants of motion $(E,L,Q)$ can equivalently be expressed in terms of these parameters by solving Eq.~\eqref{eq:circularity-condition} together with
\begin{equation}
\Theta(\theta_{\text{min}})=0\,,
\end{equation} which defines the inclination angle $I$ as in Refs.~\cite{Drasco:2005kz,Apte:2019}. 

The ISSO separates stable quasi-circular motion from unstable or plunging geodesics.   After imposing the circularity conditions in Eq.~\eqref{eq:circularity-condition}, the ISSO is identified by the additional marginal stability condition
\begin{equation}
\label{eq:stability-condition}
R''(r_{\text{ISSO}})=0\,.
\end{equation}
For fixed black hole mass and spin, and orbital inclination, this condition singles out a unique orbit within the family of circular solutions.

At the ISSO, the radial potential $R$ develops a triple root as $r_1\,,r_2\,,r_3$ in Eq.~\eqref{eq:R(r)-bounded} become degenerate, reflecting marginal stability. The radial equation for geodesics asymptoting to the ISSO can therefore be written as~\cite{Dyson:2023fws}
\begin{equation}\label{eq:plunge-eq}
   \left(\frac{\d r}{\d\lambda}\right)^2 = (1-E^2) (r_{\text{ISSO}}-r)^3 (r-r_4) 
\end{equation}where 
\begin{equation}
    r_4 = \frac{a^2 Q}{(1-{E}^2)r_{\text{ISSO}}^3}
\end{equation}
 is the remaining real root. For equatorial orbits or for a Schwarzschild background this root vanishes since  $r_4\propto Q\,,a$.

\begin{figure}
    \centering
    \includegraphics[width=\linewidth]{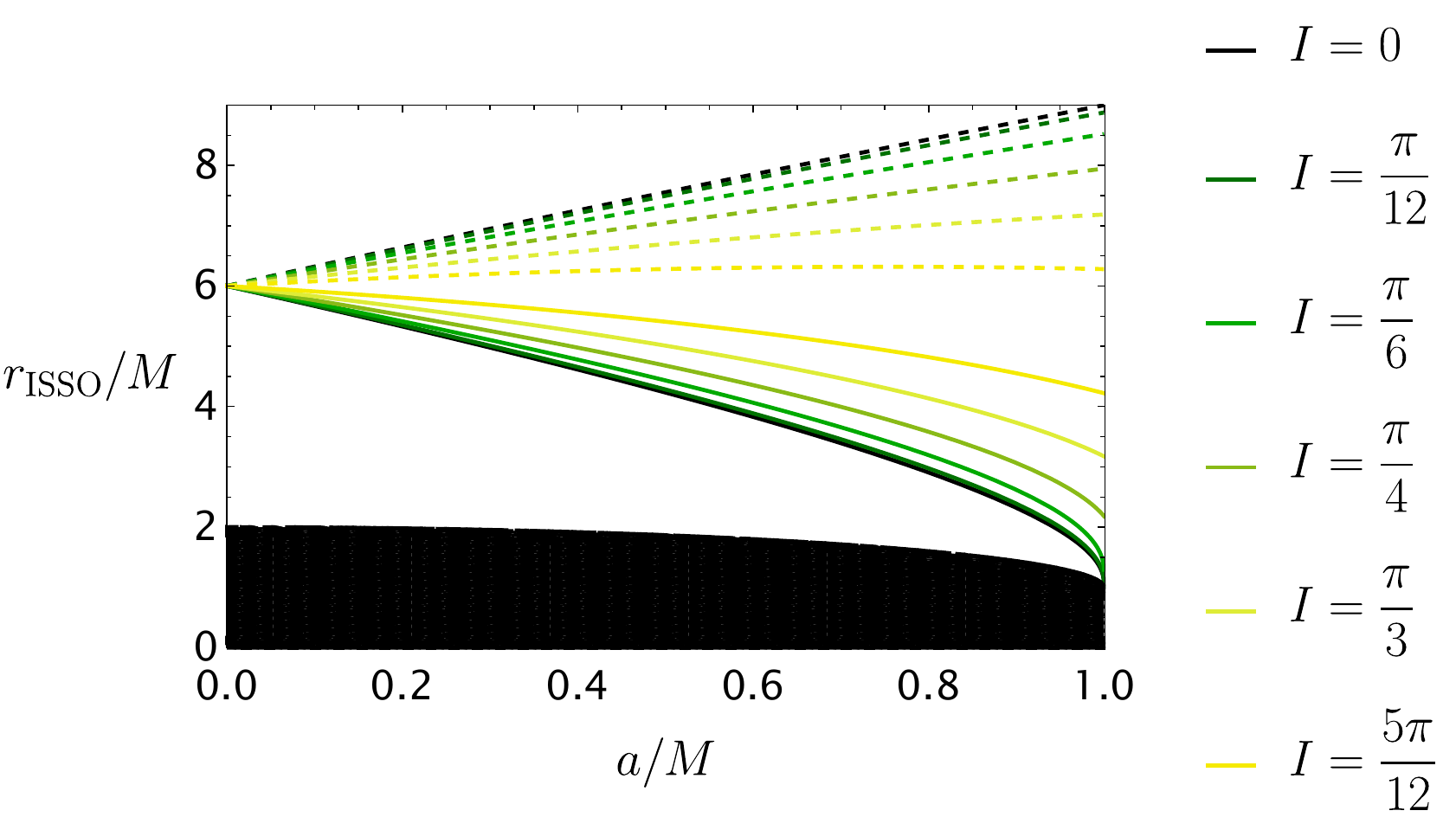}
    \captionsetup{justification=justified,singlelinecheck=false}
    \caption{Radial location of the ISSO as a function of the black hole spin  for different inclinations between the equatorial plane $I=0$, and the rotation axis $I=\pi/2$. The solid lines represent the ISSO location for prograde orbits (with $I$ as in the label and $L>0$), while the dashed lines for retrograde orbits (with $I= \pi - I_{\rm label}$ and $L<0$). The black hole region has been shaded in black.  }
    \label{fig:risco-Kerr-I}
\end{figure}

\begin{figure}
    \centering
    \includegraphics[width=\linewidth]{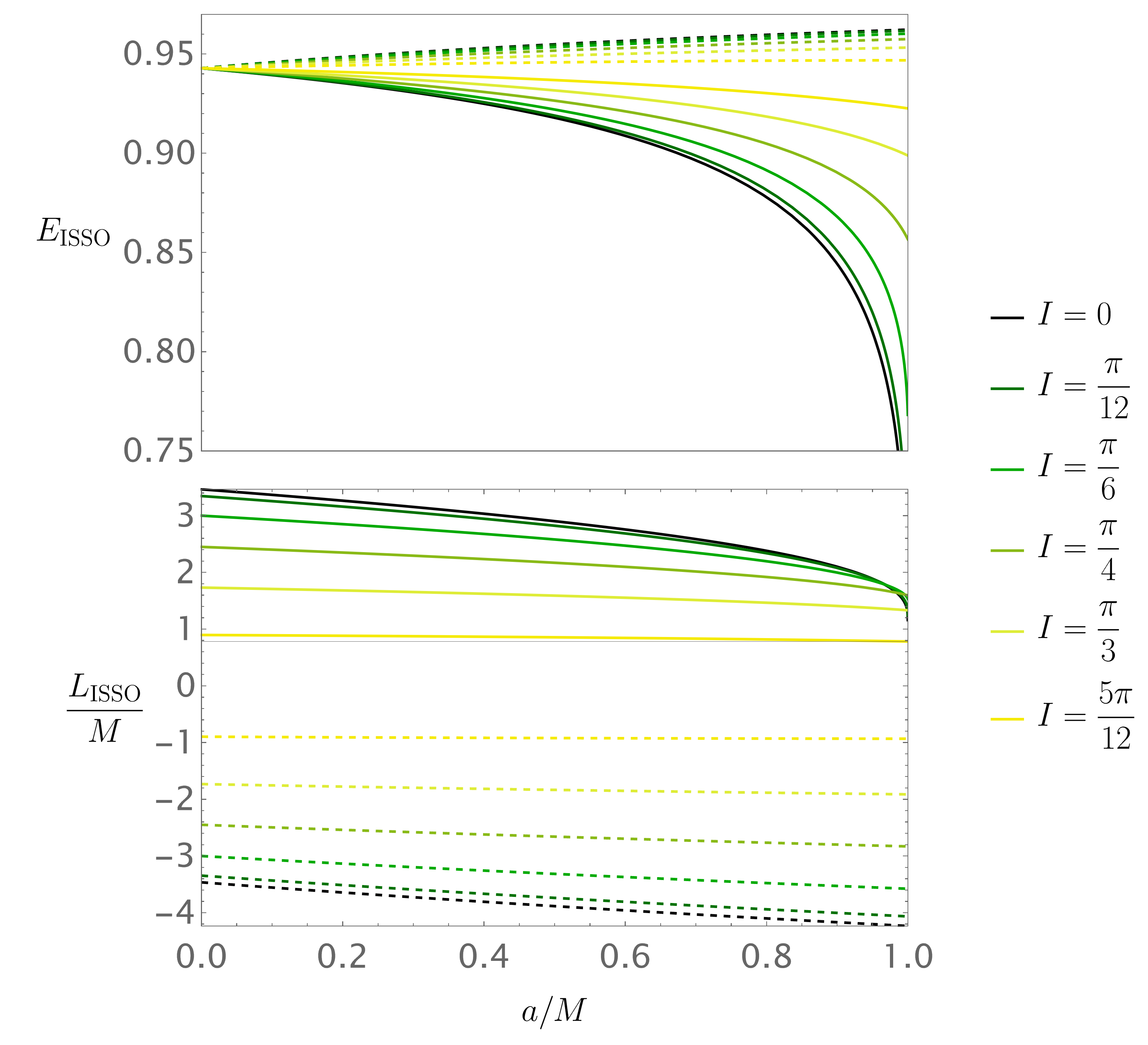}
    \caption{Energy and angular momentum at the ISSO location as a function of the black hole spin for different inclinations between the equatorial plane $I=0$, and the rotation axis $I=\pi/2$. The solid lines represent the energy and angular momentum for prograde orbits, while the dashed lines for retrograde orbits. }
    \label{fig:Eisco-Kerr-I}
\end{figure}

Solving Eqs.~\eqref{eq:circularity-condition} and~\eqref{eq:stability-condition} analytically is cumbersome, and we refer to Appendix B in Ref.~\cite{Schmidt_2002} for explicit expressions. Instead, we display the ISSO location, energy, and angular momentum for prograde (solid) and retrograde (dashed) orbits as functions of the black hole spin and inclination in Figs.~\ref{fig:risco-Kerr-I}–\ref{fig:Eisco-Kerr-I}.

In certain limits these expressions simplify greatly, and we show them for completeness. 
Introducing the dimensionless spin $\chi=a/M$ and $v=\sqrt{M/r_o}$, the Schwarzschild ($a=0$) ISCO quantities are 
\begin{subequations}   
\begin{align}
    r_{\text{ISCO}}^{a=0}/M &= 6\,,\\
    E_{\text{ISCO}}^{a=0} &= \frac{1-2v^2}{\sqrt{1-3v^2}} \,,\\
    L_{z,\text{ISCO}}^{a=0} &= \cos I \frac{M}{v\sqrt{1-3v^2}}\,,\\
    Q_{\text{ISCO}}^{a=0} &= \sin^2I \frac{M^2}{v^2 (1-3v^2)} \,.
    \end{align}
\end{subequations} For equatorial orbits in Kerr ($I=0$), these quantities read
\begin{subequations}   
\begin{align}\label{eq:risco-I=0}
    r_{\text{ISCO}}^{I=0}/M &=  3+Z_2 \mp \sqrt{(3-Z_1)(3+Z_1+2Z_2)}\,,\\
    E_{\text{ISCO}}^{I=0} &= \frac{1-2v^2 \pm \chi v^3}{\sqrt{1-3v^2\pm \chi v^3}}\,,\\
    L_{z,\text{ISCO}}^{I=0} &= \pm \frac{M}{v} \frac{1\mp 2 \chi v^3 + \chi^2 v^4}{\sqrt{1-3v^2 \pm 2 \chi v^3}}\,,\\
    Q_{\text{ISCO}}^{I=0} &= 0\,,
    \end{align}
\end{subequations} with
\begin{subequations}
    \begin{align}
    Z_1 &\equiv 1+(1-\chi^2)^{1/3} \left[(1+\chi)^{1/3}+(1-\chi)^{1/3}\right]\,,\\
    Z_2 &\equiv \sqrt{3\chi^2 +Z_1^2}\,.
\end{align}
\end{subequations}

\subsection{The transition-to-plunge for inclined orbits in Kerr}
\label{subseec:transition-to-plunge}

The derivation of the differential equation governing the radial motion across the transition to plunge was first presented by Ori and Thorne~\cite{Ori:2000zn} and Buonanno and Damour~\cite{Buonanno:2000ef} for quasi-circular, equatorial orbits in Kerr, and later generalized to inclined orbits by Apte and Hughes~\cite{Apte:2019}. Here we review this derivation for quasi-circular inclined motion, focusing on the radial dynamics near the ISSO.

The description of the transition relies on three key ingredients:
\begin{enumerate}
    \item Radiation reaction drives a slow evolution of the geodesic constants of motion compared to the orbital timescale. A multiscale description therefore applies: the orbital motion is characterized by the fast timescale $T_o$, of the order of the orbital period, while the constants of motion evolve on the radiation-reaction timescale $T_{\text{RR}}$. Their ratio scales as
    \begin{equation*}
        \frac{T_o}{T_{\text{RR}}} \sim \eta
    \end{equation*} where $\eta = \frac{\mu}{M}$ is the mass-ratio, which we take to be $\eta\ll1$ for an EMRI. 
    \item The slow evolution of the constants of motion can be viewed as a slow deformation of the effective radial potential~\cite{Ori:2000zn}. The transition occurs when the local minimum and maximum of this potential merge into an inflection point, i.e., when the ISSO conditions are satisfied.
    \item To leading order in the mass ratio, the conservative part of the self-force, which shifts the ISSO location, can be neglected~\cite{Ori:2000zn,Apte:2019}. The transition is thus modeled as geodesic motion in a slowly evolving background driven by the dissipative self-force.
\end{enumerate}

The transition regime has since been studied systematically beyond leading order in Schwarzschild{~\cite{Buonanno:2000ef,Kuchler:2025hwx,Compère_2020-Painleve}}, and recently extended to equatorial Kerr orbits~\cite{Kuchler:2024esj,Compere:2021iwh,Honet:2025dho}. Here, however, we restrict attention to the leading-order correction to geodesic motion.

Following~\cite{Apte:2019}, we start the derivation of the differential equation governing the radial motion during the transition by rewriting the radial geodesic equation in Eq.~\eqref{eq:geodesic-Kerr-Mino} as a second-order differential equation, 
\begin{equation}\label{eq:radial-geodesic-2nd-order}
\begin{split}
    & \frac{\d^2r}{\d\lambda^2} = \frac{1}{2}\frac{\partial R}{\partial r}\\
     &+\frac{1}{2\left(\frac{\d r}{\d\lambda}\right)}\left(\left(\frac{\partial R}{\partial E}\right)  \frac{\d E}{\d\lambda}+\left(\frac{\partial R}{\partial L}\right)  \frac{\d L}{\d\lambda}+\left(\frac{\partial R}{\partial Q}\right)  \frac{\d Q}{\d\lambda}\right)\,.
\end{split}
\end{equation} where $R = R(r\,,\mathcal{C}(\lambda))$, with $\mathcal{C} = (E\,,L\,,Q)$ now slowly evolving due to radiation reaction, and $r$ labeling a particular orbital radius. 

Near the ISSO, the radial motion and the evolution of the constants $\mathcal{C}$ both scale with the mass-ratio $\eta$, but the two lines in Eq.~\eqref{eq:radial-geodesic-2nd-order} scale differently. The first line is of order $\mathrm{O}(\eta)$ around the ISSO since we consider circular orbits. The second scales with $\mathrm{O}(1)$ since both $\frac{\d\mathcal{C}}{\d\lambda}\sim \frac{\d r}{\d\lambda}\sim \mathrm{O}(\eta)$~\cite{Apte:2019}. 
Requiring consistency order-by-order in $\eta$ yields two equations
\begin{subequations}\label{eq:second-order-transition}
    \begin{align}\label{eq:second-order-transition:1}
    \frac{\d^2r}{\d\lambda^2} =& \frac{1}{2}\frac{\partial R(r\,,\mathcal{C}(\lambda))}{\partial r}\\\label{eq:second-order-transition:2}
    0=&\left(\frac{\partial R}{\partial E}\right)  \frac{\d E}{\d\lambda}+\left(\frac{\partial R}{\partial L}\right)  \frac{\d L}{\d\lambda}+\left(\frac{\partial R}{\partial Q}\right)  \frac{\d Q}{\d\lambda}
\end{align}
\end{subequations}
 where the quantities in parentheses are constants evaluated at the ISSO.

 The first equation, expanded around the ISSO, yields the transition equation. The second equation is the quasi-circularity condition, first identified by Kesden~\cite{Kesden:2011ma} for equatorial motion and generalized to inclined orbits in~\cite{Apte:2019}.  In the following, we review the argument in Ref.~\cite{Kesden:2011ma}.

For circular equatorial orbits, the evolution of the energy and angular momentum are related so that circular orbits evolve to circular orbits. In fact, in {Ref.}~\cite{Ori:2000zn}, Ori and Thorne took the rate of change of energy and angular momentum to be proportional, satisfying
\begin{equation}
    \Omega^{-1}\frac{\d E}{\d\tau} - \frac{\d L}{\d\tau}=0
\end{equation} with $\tau$ proper time, and $\Omega$ the orbital frequency. Notice that this condition is equivalent to Eq.~\eqref{eq:second-order-transition:2} being satisfied, since $\Omega^{-1} = -\partial_E R/\partial_L R$, and it is valid to leading order. 

However, as shown by Kesden~\cite{Kesden:2011ma}, this condition leads to an inconsistency in the normalization of the four-velocity if the next-to-leading order correction is not accounted for, which is manifest in the first-order version of Eq.~\eqref{eq:second-order-transition:1}.  The correct condition to next-to-leading order is 
\begin{equation}\label{eq:Kesden-condition}
    \Omega^{-1} \frac{\d E}{\d\tau} -\frac{\d L}{\d\tau} \propto\eta^{4/5} (r-r_{\text{ISCO}})\,,
\end{equation} i.e., the proportionality between the rate of change of energy and angular momentum holds only up to higher-order corrections in the mass-ratio~\footnote{This is why the orbits are called ``quasi''-circular.}. 

For the derivation in the case of inclined orbits, we follow~\cite{Apte:2019} and impose directly the generalized quasi-circularity condition in Eq.~\eqref{eq:second-order-transition:2} to leading order in $\eta$. This is enough for this work since we restrict ourselves to the study of the leading order transition. For higher order refinements of the transition dynamics one should take into account a correction of type~\eqref{eq:Kesden-condition} (see e.g.,~\cite{Compere:2021zfj}). 
To derive the transition equation, we  expand Eq.~\eqref{eq:second-order-transition:2} around the ISSO. We introduce
\begin{subequations}\label{eq:expansion-variables-transition}
\begin{align}\label{eq:expansion-variables-transition:1}
    x(\lambda) &= r(\lambda)-r_{\text{ISSO}}\,,\\\label{eq:expansion-variables-transition:2}
        \delta E(\lambda) &= E(\lambda)-E_{\text{ISSO}} \,,\\\label{eq:expansion-variables-transition:3}
        \delta L(\lambda) &= L(\lambda)-L_{\text{ISSO}} \,.\\\label{eq:expansion-variables-transition:4}
        \delta Q(\lambda) &= Q(\lambda)-Q_{\text{ISSO}} \,.
\end{align}     
\end{subequations} 

Using the ISSO conditions in  Eqs.~\eqref{eq:circularity-condition} and~\eqref{eq:stability-condition}, together with the expansion in Eq.~\eqref{eq:expansion-variables-transition} the radial potential can be expanded as 
\begin{equation}
\begin{aligned}
R= & \frac{1}{6}\left(\frac{\partial^3 R}{\partial r^3}\right) x^3 \\
& +\left(\frac{\partial^2 R}{\partial r \partial E} \delta E+\frac{\partial^2 R}{\partial r \partial L} \delta L+\frac{\partial^2 R}{\partial r \partial Q} \delta Q\right) x \\
& + \text { terms independent of } x\,.
\end{aligned} 
\end{equation}
The evolution of the energy, angular momentum, and Carter constant are purely driven by radiation reaction, and computing them exactly would require knowledge of the self-force. However, the transition takes a short time compared to the radiation-reaction timescale. In particular~\cite{Apte:2019},
\begin{equation}
    \lambda-\lambda_{\text{ISSO}} \sim \eta^{-1/5}\,,
\end{equation}
and hence we can approximate the unknown functions $\delta\mathcal{C}$ by a Taylor expansion around the ISSO, i.e., 
\begin{equation}\label{eq:delta-c}
    \delta\mathcal{C} = \eta \sum_{i=1}^\infty \eta^{\frac{i-1}{5}} \left(\frac{\d^{i} \mathcal{C}}{\d\lambda^{i}}\right)  (\lambda-\lambda_{\text{ISSO}})^i \,.
\end{equation} where we assume that higher powers in $(\lambda-\lambda_{\text{ISSO}})$ are suppressed by  higher powers in the mass-ratio. This assumption is revisited in Sec.~\ref{subsec:generalization}.

Since we wish to derive the transition equation to leading order, the evolution of the constants of motion is linear in $\lambda-\lambda_{\text{ISSO}}$
\begin{equation}\label{eq:C-evolution}
    \delta\mathcal{C} = \eta \, \kappa_\mathcal{C} (\lambda-\lambda_{\text{ISSO}})+\mathrm{O}(\eta^{6/5})\,,
\end{equation} where the constants $\kappa_\mathcal{C}$ can be read off from Eq.~\eqref{eq:delta-c}. 
Introducing the constants 
\begin{equation}\label{eq:A-B-def}
\begin{gathered}
A \equiv-\frac{1}{4}\left(\frac{\partial^3 R}{\partial r^3}\right), \\
B\equiv \frac{1}{2}\left(\frac{\partial^2 R}{\partial r \partial E} \kappa_E+\frac{\partial^2 R}{\partial r \partial L} \kappa_{L}+\frac{\partial^2 R}{\partial r \partial Q} \kappa_Q\right) ,
\end{gathered}
\end{equation} allows us to rewrite Eq.~\eqref{eq:second-order-transition:1} in the simplified form
\begin{equation}\label{eq:Painleve-eta}
    \frac{\d^2 x}{\d\lambda^2} = -A x^2 +B\eta (\lambda-\lambda_{\text{ISSO}})\,.
\end{equation}
Finally, rescaling the relevant quantities by appropriate powers of the mass-ratio through the transformation
\begin{equation}\label{eq:transformation-X_T}
    \begin{gathered}
x  \equiv \eta^{2 / 5} B^{2 / 5} A^{-3 / 5} X, \\
\left(\lambda-\lambda_{\mathrm{ISSO}}\right) \equiv  - \eta^{-1 / 5}(A B)^{-1 / 5} T ,
\end{gathered}
\end{equation} 
yields the Painlev\'e I differential equation
\begin{equation}\label{eq:PainleveI}
    \frac{\d^2 X}{\d T^2} = -X^2-T
\end{equation} as first identified in Ref.~\cite{Compère_2020-Painleve}.

\section{
An approximate exact solution to the Painlev\'e I differential equation}
\label{sec:Ori-Thorne-Kesden}
{In the seminal works of Buonanno and Damour~\cite{Buonanno:2000ef} and Ori and Thorne~\cite{Ori:2000zn}, it was noted that Eq.~\eqref{eq:PainleveI} does not admit a solution in terms of elementary or special functions.} The authors therefore solved the equation numerically, imposing the asymptotic boundary condition
\begin{equation}
\label{eq:bdy-adiabatic}
X(T) \sim \sqrt{-T},
\qquad T \to -\infty,
\end{equation}
together with the consistent condition on its derivative $\dot{X}(T) \sim -\frac{1}{2}(-T)^{-1/2}$ as $T\to -\infty$. This boundary condition ensures consistency with the adiabatic inspiral regime at early times, where $\ddot{X}\to0$ (with $\dot{\quad} = \d/\d T$).  

At late times, as the system approaches plunge, the variable $T$ varies slowly and Eq.~\eqref{eq:PainleveI} reduces locally to
\begin{equation}
\ddot X \approx -X^2,
\end{equation}
whose solutions develop a second-order pole,
\begin{equation}
\label{eq:bdy-plunge}
X(T) \sim \frac{-6}{(T_{\text{plunge}}-T)^2},
\end{equation} with $T_{\text{plunge}}\approx 3.412$, fixed after imposing the asymptotic adiabatic boundary condition in Eq.~\eqref{eq:bdy-adiabatic}. The appearance of this pole signals the breakdown of the transition approximation and the onset of geodesic plunge, {as noted by Buonanno and Damour~\cite{Buonanno:2000ef} and Ori and Thorne~\cite{Ori:2000zn}}.

It was later recognized in Ref.~\cite{Compère_2020-Painleve} that Eq.~\eqref{eq:PainleveI} is precisely the first Painlevé equation, { or Painlevé I equation,} one of the six nonlinear differential equations identified by Painlevé and collaborators that defined transcendental functions not known before. Although their solutions cannot be expressed in terms of known special functions, they are nevertheless analytic functions with highly constrained singularity structure. These solutions are known as the Painlevé transcendents. 

The first Painlevé equation admits a two-parameter family of solutions. However, the asymptotic condition in Eq.~\eqref{eq:bdy-adiabatic} uniquely selects a distinguished solution: the \emph{tritronquée} solution. This special solution is characterized by the absence of poles in a maximal sector of the complex plane containing the negative real axis, and by a discrete sequence of poles along the positive real axis. In particular, the pole at $T_{\text{plunge}}$ corresponds precisely to the first real singularity of the tritronquée solution.

{ In \cite{Jaramillo:2022oqn}, the Ori–Thorne numerical solution was explicitly identified with Boutroux's} tritronquée solution of the Painlevé I equation. { In this work we develop and extend this identification, crucially providing a first-principles analysis. Such} identification is not merely formal: it allows us to exploit the extensive mathematical results available for this special function. In particular, we employ the explicit analytic representation constructed in Ref.~\cite{ADALITANVEER20163843}, which provides an approximate closed-form expression with rigorously bounded errors for both the solution and its derivatives.

We then compare this explicit approximation with direct numerical integration of Eq.~\eqref{eq:PainleveI}. Because the analytic representation comes with controlled error bounds, it provides improved accuracy near the plunge, where the errors grow significantly for a ``naive'' numerical integration. We present the construction of the tritronquée solution in Sec.~\ref{sec:tritronque-sol}, and analyze its performance relative to numerical methods in Sec.~\ref{sec:performance}.

\subsection{The tritronqu\'ee solution of the Painlev\'e I {transcendent}}
\label{sec:tritronque-sol}

Singularities of differential equations can be divided into two types: \emph{fixed} singularities, whose location is determined solely by the equation itself, and \emph{movable} singularities, whose location depends on the initial data. Linear second-order differential equations admit only fixed singularities. Nonlinear equations, in contrast, may also develop movable singularities. Among the movable singularities that can arise, poles are the mildest. A distinguished subclass of nonlinear equations allows no movable singularities other than poles; consequently, their solutions are meromorphic throughout the complex plane. Such equations are said to possess the \emph{Painlev\'e property}.

At the beginning of the twentieth century, Painlevé{, with complements by Gambier and Fuchs} classified all second-order differential equations of the form
\begin{equation}\label{eq:Painleve-classification}
y'' = \mathcal{F}(y',y,x),
\end{equation}
with $\mathcal{F}$ rational in $(y',y)$, that possess the Painlevé property. Besides equations reducible to known special functions, six irreducible equations were found, defining new transcendental functions known as the Painlevé transcendents. The simplest of these is the Painlevé I equation, usually written as~\cite{NIST:DLMF}
\begin{equation}
\label{eq:Painleve-I-main}
y'' = 6y^2 + x .
\end{equation}

The equation governing the transition-to-plunge, Eq.~\eqref{eq:PainleveI}, differs from Eq.~\eqref{eq:Painleve-I-main} only by trivial rescalings of the dependent and independent variables and is therefore equivalent to Painlevé I~\footnote{A class of transformations which leaves invariant the singularity structure in Eq.~\eqref{eq:Painleve-classification} is the homographic group (also called M\"{o}bius group $PSL(2\,,\mathbb{C})$)
\begin{equation*}
    (y\,,x) \mapsto (Y,X)\,, \quad y(x) = \frac{\alpha(x) Y(x) +\beta (x)}{\gamma(x) Y(x) +\delta (x)}\,,\quad X=\xi(x)
\end{equation*} 
where $(\alpha\,,\beta\,,\gamma\,,\delta\,,\xi)$  are functions such that $\alpha\delta-\beta \gamma\neq 0$. Rescalings by a constant of both the dependent and independent variables are trivial representations of the homographic group.}. All solutions to the Painlev\'e I equation are meromorphic in the complex plane, with an essential singularity at infinity.

A remarkable feature of the Painlevé transcendents is the organization of its poles in the complex plane. For Painlev\'e I, Boutroux~\cite{Boutroux_1914} showed that near infinity the complex plane is divided into five Stoke sectors separated by the rays $\arg x = 2\pi n/5$ with $n=0,\pm1,\pm2$~\footnote{When you have an asymptotic expansion of a function valid as $|x| \to \infty$, the expansion can behave very differently depending on which direction you go to infinity. Crossing one of the rays (called a
Stokes line) can cause subdominant terms in the asymptotic expansion to suddenly ``switch on'';  this is the Stokes phenomenon. Within each sector, the asymptotic behavior is uniform, but it can change discontinuously as you cross a ray.
}. In general, in each Stokes sector the solution has a well-defined asymptotic expansion as $|x| \to \infty$. For Painlev\'e I, the generic solution has poles accumulating toward all five rays, so it does not have a clean asymptotic description in any sector. 
However, special one-parameter families of solutions, called \emph{tronqu\'ee solutions}, are asymptotically pole-free along one of these five rays. For any such tronqu\'ee solution, the two adjacent sectors to the special ray are also pole free, and hence an asymptotic solution can be built in these two sectors. 
If a solution is truncated along two adjacent rays, it is in fact truncated along three consecutive rays and asymptotically pole-free in four consecutive sectors; such a solution is known as the 
\emph{tritronqu\'ee solution}. There exist five such solutions, related by the fivefold rotational symmetry of Painlevé I $y\mapsto e^{4i k\pi/5} y(e^{2\pi i k/5 } x)$ for $k=0\,,...\,,4$~\cite{Dubrovin-conj:2012}.
In other words, the tronquée and tritronquée solutions are special precisely because they are pole-free in some sectors, which is what allows a valid asymptotic expansion to be constructed there.

We now return to the physical boundary condition relevant for the transition regime. 
Imposing the adiabatic boundary condition at early times in Eq.~\eqref{eq:bdy-adiabatic} fixes the two integration constants of the second-order equation and therefore uniquely determines the solution. Remarkably, the asymptotic condition that the orbit be quasi-circular at early times (see Eq.~\eqref{eq:circularity-condition}) selects precisely the real tritronquée solution aligned with the negative real axis (using the convention of Eq.~\eqref{eq:PainleveI}). This solution is pole-free along the whole real negative axis, consistent with the fact that we can match the transition solution to the adiabatic inspiral regime as $T\to-\infty$, but develops a pole at a finite positive value, which Ori and Thorne called $T_{\text{plunge}}$; we will refer to it as $T_{\text{pole}}$.

Since the tritronqu\'ee solution of the Painlev\'e I differential equation is a real-analytic function with isolated pole-type singularities (a function in the sense of Painlev\'e), one may construct an asymptotic expansion of the form 
\begin{equation}\label{eq:asumptotic-X0}
    X(T)\sim \sqrt{-T} \sum_{k=0}^\infty  \frac{a_k}{(-T)^{5k/2}}\,,\quad T\to-\infty
\end{equation} with coefficients $a_k$ determined recursively. However, this series is asymptotic rather than convergent and does not remain accurate up to the pole. Similarly, numerical methods solving Eq.~\eqref{eq:PainleveI} with asymptotic boundary conditions (for some large but finite $T\in \mathbf{R}^-$) lose precision close to the pole. 

An alternative approach is to construct an explicit approximate representation 
\begin{equation}
X(T) = X_0(T) + E(T),
\end{equation}
where $X(T)$ is the solution to Eq.~\eqref{eq:PainleveI}, $X_0(T)$ is the approximation given in closed form and the error $E(T)$ is rigorously bounded over the whole transition domain. 
In Ref.~\cite{ADALITANVEER20163843}, Adali and Tanveer constructed such an approximation using an asymptotic-matching procedure and derived explicit error bounds for both the solution and its derivative,
\begin{equation}\label{eq:error-bounds}
    |E(T)| \leq 6.89\times 10^{-5}\,, \quad |\dot{E}(T)|\leq 2.38\times 10^{-4}\,,
\end{equation} which, as we will see in Sec.~\ref{sec:performance} are comparable to existing numerical methods. They further obtained an accurate approximation for the pole location,
\begin{equation}\label{eq:T0}
T_0 = \frac{770766}{323285} 6^{1/5} \approx 3.41167 ,
\end{equation}
with rigorous bound
\begin{equation}
|T_{\text{pole}} - T_0| \leq 5.9\times 10^{-6}\,.
\end{equation}

We now present their approximate solution $X_0(T)$, omitting the details of the derivation. To ensure a maximal error bound as in Eq.~\eqref{eq:error-bounds}, the solution presented in Ref.~\cite{ADALITANVEER20163843} is piecewise, divided in three regions
\begin{subequations}
\begin{align}
    R_1 &= (-\infty, -6^{1/5} \frac{11}{2}]\approx (-\infty\,,-7.87]\,,\\
    R_2 & = (-6^{1/5} \frac{11}{2}\,,T_0-6^{1/5}\frac{7}{10}] \approx (-7.87\,,2.41]\,,\\
    R_3 & = \{T\in\mathbb{C} : |T-T_0| =6^{1/5}  \frac{7}{10}\,, T \neq T_0-6^{1/5}  \frac{7}{10}\}\,.
\end{align} 
\end{subequations}
Note that the region $R_3$ extends to the complex plane. 
The error on the solution and its first derivative on the joining points across regions is of the order of $10^{-14}$, so way below the error bounds. 

The approximate solution is given by~\cite{ADALITANVEER20163843}
\begin{widetext}
\begin{equation}\label{eq:approx-exact-sol-X0}
    X_0(T)  = \left\{\begin{matrix}
        \sqrt{-T} \left[1+\frac{1}{8} (-T)^{-5/2} -\frac{49}{128} (-T)^{-5}+\frac{1225}{256} (-T)^{-15/2} +\widetilde{w}_0(T)\right] & \text{on } R_1\\
        -\frac{6}{(T-T_0)^2} +6^{3/5}P_2 (T) & \text{on } R_2\\
        -\frac{6}{(T-T_0)^2} +6^{1/5}(T-T_0)^2 P_3((T_0-T)/6^{1/5}) & \text{on } R_3
    \end{matrix}
    \right.
\end{equation} 
\end{widetext}
with 
\begin{equation}
\label{eq:wo}
    \widetilde{w}_0(T) = \re \left[\int_0^\infty e^{-s\frac{4\sqrt{2}}{5}(-T)^{5/4}} \mathcal{W}_0 \left( \tfrac{(-T)^{5/4}}{6^{1/4}}\,,s \right) \d s\right]\,,
\end{equation} and the functions $P_2(T)$ and $P_3(T)$ 
\begin{subequations}
 \begin{align}\label{eq:P2-P3}
    P_2(T) &=\sum_{k=0}^{22} b_k \left(-\frac{T-T_0/2+\frac{31}{10} 6^{1/5}}{T_0/2 +\frac{12}{5} 6^{1/5}}\right)^k \\
    P_3(T) &= \sum_{k=0}^{17} c_k T^k
\end{align}   
\end{subequations}
 are finite polynomials in $T$. 
The numerical values of the coefficients $b_k$ and $c_k$ and the function $\mathcal{W}_0$ are given in App.~\ref{sec:extra-approximate-sol}. 

The first four terms of the expression in $R_1$ have appeared previously in the literature (see e.g.~\cite{Apte:2019}), as they coincide with the large-$|T|$ asymptotic expansion in Eq.~\eqref{eq:asumptotic-X0}. The significance of the representation in Eq.~\eqref{eq:approx-exact-sol-X0}, however, is that it is not merely an asymptotic expansion valid as $T\to -\infty$, but a uniformly valid approximation across the transition domain, with rigorous error bounds.

Strictly speaking, this validity does not extend all the way to the pole as $R_3$ does not cover all the way to the pole location $T_0$ along the real line.  Let's denote the region extending to a distance  $\epsilon$ from the pole along the real line by  
\begin{equation}
    R_4 = \left(-6^{1/5}\frac{7}{10},\, T_0-\epsilon\right)\,,
\end{equation}
for $\epsilon\ll1$. The region $R_4$ is not part of the solution presented in Eq.~\eqref{eq:approx-exact-sol-X0}, because if we analytically continue the solution in $R_3$ to $R_4$ the error bound \eqref{eq:error-bounds} established in Ref.~\cite{ADALITANVEER20163843} is no longer guaranteed to hold. 
Nevertheless, as we will show below, the extension of the solution in $R_3$ into $R_4$ continues to yield errors well below the bounds in Eq.~\eqref{eq:error-bounds} for $\epsilon \lesssim10^{-3}$, so we will extend the solution in $R_3$ all the way to $T_0-\epsilon$. 

\subsection{Performance of the exact approximate solution}
\label{sec:performance}

In this section we assess the performance of the exact approximate representation in Eq.~\eqref{eq:approx-exact-sol-X0} by comparing it with a high-precision numerical integration of Eq.~\eqref{eq:PainleveI} in \textit{Mathematica}. While numerical solutions of the Painlevé I equation can achieve excellent pointwise accuracy away from the singularity, they do not provide uniform analytic control near the pole, nor guaranteed bounds on derivatives. 
The approximate representation constructed in Ref.~\cite{ADALITANVEER20163843} provides global bounds on the error, and has therefore the potential to behave better near the pole.

For the numerical solution, we solve the Painlev\'e I equation using \textit{NDSolve} in \textit{Mathematica}, 
with \texttt{WorkingPrecision} set to 50, \texttt{AccuracyGoal} set to $\infty$, 
and the \texttt{ImplicitRungeKutta} method. These parameters were chosen to match those employed in the numerical studies of Refs.~\cite{Kuchler:2024esj,pc-Lorenzo}. 

Initial data are imposed at $T=-40$ using the asymptotic expansion
\begin{equation}
\begin{split}
     X_{\text{Asymp}}(T) = & \sqrt{-T} 
    \left[1+\frac{1}{8} (-T)^{-5/2} 
    \right.\\
    &\left.-\frac{49}{128} (-T)^{-5}  +\frac{1225}{256} (-T)^{-15/2} \right],
\end{split}
\end{equation}
together with its derivative evaluated at the same point. 

We have verified that varying the numerical parameters in \textit{NDSolve}---for example by increasing the working precision, changing the integration method, or imposing the initial data further in the asymptotic regime at $T=-100$---does not lead to any significant improvement in the numerical solution presented here.
Although both the numerical and analytical treatments could, in principle, be further optimized, the comparison below shows that the analytical approximation already achieves a level of accuracy comparable to that of a reasonable  numerical solution. In this sense, the analytical expression provides a practical alternative to direct numerical integration while retaining rigorous error control.

\begin{figure}
    \centering
    \includegraphics[width=\linewidth]{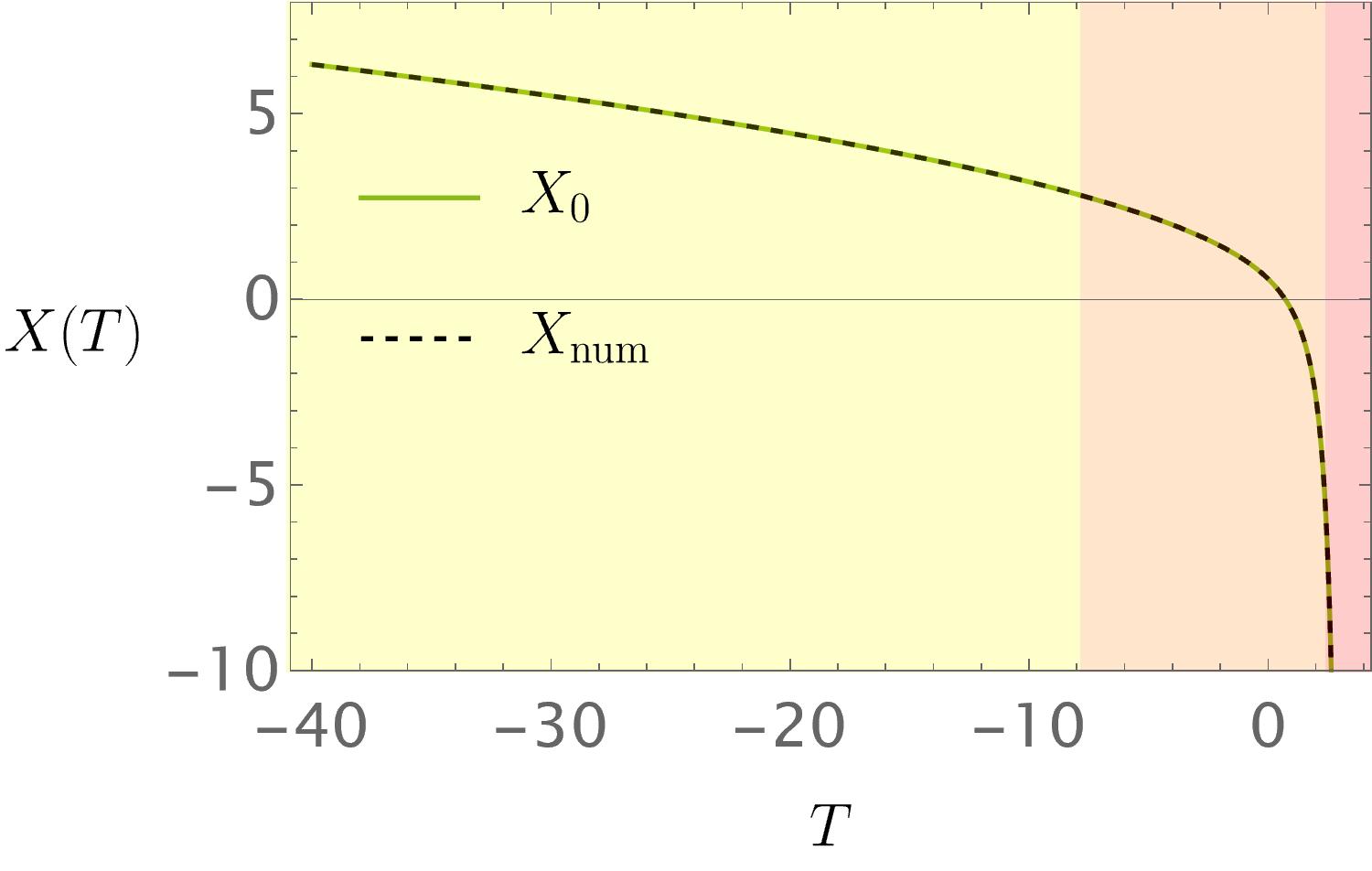}
    \caption{
    Comparison between the approximate tritronqu\'ee solution $X_0(T)$ 
and the numerical solution $X_{\text{num}}(T)$ of the Painlev\'e I equation 
as functions of $T$. The two curves are visually indistinguishable on this scale 
throughout the transition regime up to $T_0-\epsilon$. We colored the regions $R_1$ (yellow), $R_2$ (orange), and $R_3\cup R_4$ (red) of the approximate solution. }
    \label{fig:X(T)}
\end{figure}

In Fig.~\ref{fig:X(T)}, we display the approximate solution $X_0$ and the numerical solution $X_{\text{num}}$ up to $T_0-\epsilon$, where throughout this section we take $\epsilon=10^{-3}$. The regions $R_1$, $R_2$, and $R_3\cup R_4$ are colored  yellow, orange, and red, respectively.  
The two curves are visually indistinguishable on this scale. 

\begin{figure}
    \centering
    \includegraphics[width=\linewidth]{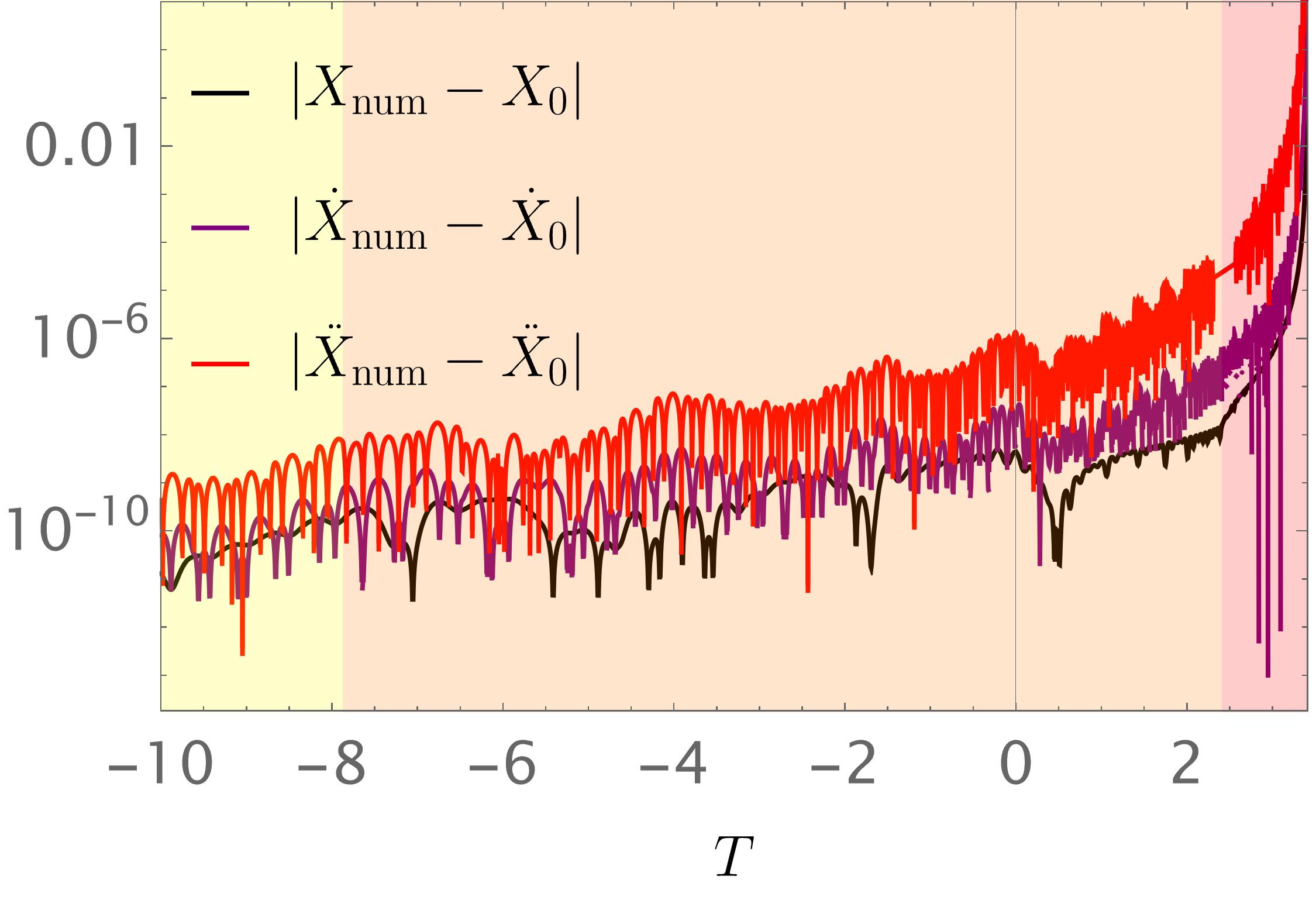}
    \caption{
   Absolute difference between the approximate solution $X_0$ and the 
numerical solution $X_{\text{num}}$ (black), together with the absolute 
difference of their first (purple) and second (red) derivatives, 
plotted on a logarithmic scale. While the functions themselves agree 
to high accuracy across the transition, discrepancies increase for 
higher derivatives as the pole is approached. 
}
    \label{fig:X_Xp_Xpp}
\end{figure}

To quantify the agreement, Fig.~\ref{fig:X_Xp_Xpp} shows the absolute difference 
between $X_0$ and $X_{\text{num}}$, as well as the difference between their first and second derivatives, plotted on a logarithmic scale. 
The discrepancy between the functions themselves remains extremely small throughout most of the transition and increases only as $T\to T_0$. 

The differences in the derivatives are more pronounced. In particular, the discrepancy in the second derivative reaches $\mathrm{O}(10^{-3})$ around $T\approx 3$, still well before the pole. This behavior reflects the well-known sensitivity of numerical differentiation near singularities.

While Fig.~\ref{fig:X_Xp_Xpp} measures the mutual agreement of the two solutions, it does not indicate which one more accurately satisfies the differential equation. 
To assess this, we introduce the following two residuals
\begin{subequations}
\begin{align}\label{eq:Delta}
    \Delta(T) &= \ddot{X} + X^2 + T,\\ \label{eq:Delta:tilde}
    \widetilde{\Delta} (T) &=  -(X^\prime)^2-\frac{2}{3} X^3 -2T X+Y 
\end{align}    
\end{subequations} with $Y$ satisfying $\frac{\d Y}{\d T} = 2X$. The two residuals are defined as the transition differential equation and its first integral (following Ref.~\cite{Kesden:2011ma}), using the rescaled variables in Eq.~\eqref{eq:transformation-X_T}.

Both residuals vanish identically for an exact solution of the Painlev\'e I equation. 
For approximate or numerical representations, $\Delta$ and $\widetilde \Delta$ provide direct diagnostics of how well the differential equation and its first integral are satisfied.

\begin{figure}
    \centering
    \begin{subfigure}{0.48\textwidth}
        \centering 
         \includegraphics[width=\linewidth]{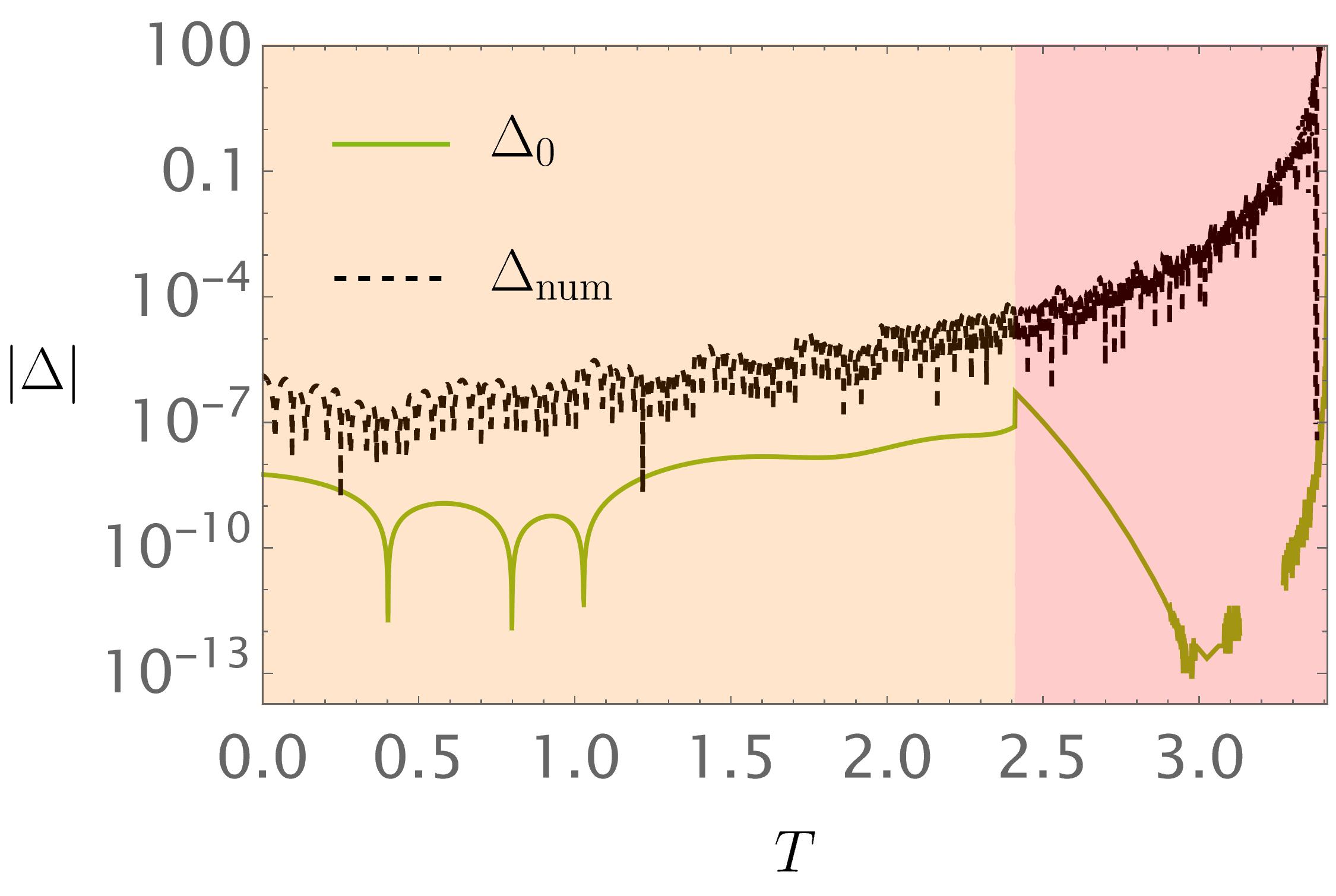}
         \caption{ Absolute value of the residual 
$\Delta(T)$ in Eq.~\eqref{eq:Delta}. }
     \label{fig:delta-residual}
    \end{subfigure}
   \hfill
   \begin{subfigure}{0.48\textwidth}
   \centering
    \includegraphics[width=\linewidth]{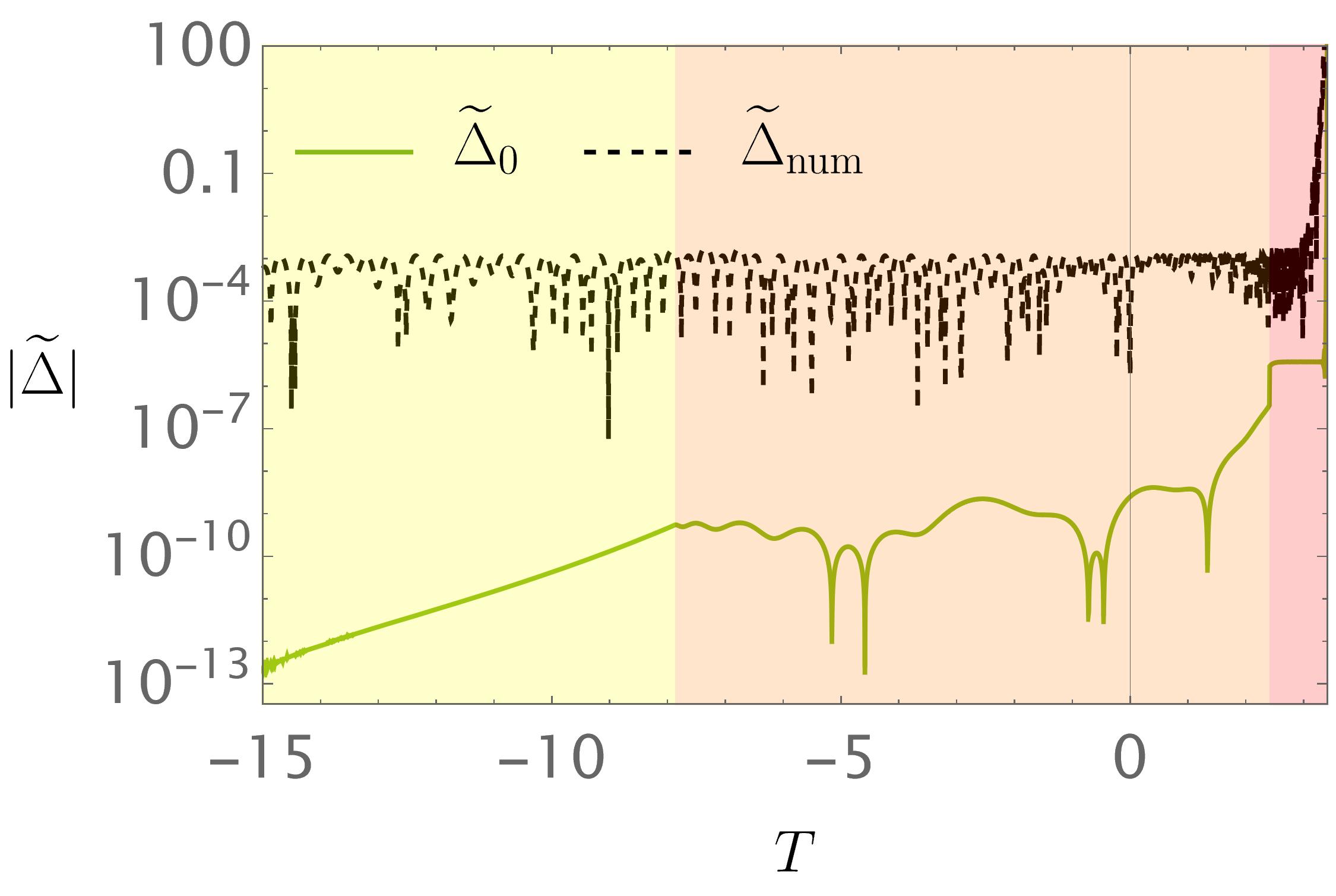}
    \caption{Absolute value of the residual 
$\widetilde{\Delta}(T)$ in Eq.~\eqref{eq:Delta:tilde}.}
    \label{fig:tilde-delta-residual}
   \end{subfigure}
    \caption{ 
   Comparison of the residuals $\Delta$ and $\widetilde{\Delta}$  
evaluated using the approximate solution $X_0$ (solid green) 
and the numerical solution $X_{\text{num}}$ (dashed black), 
shown on a logarithmic scale for $0\leq T\leq T_0-\epsilon$ and $-15\leq T\leq T_0-\epsilon$ respectively. 
The bounded residual of $X_0$ indicates uniform consistency 
with the differential equation, while the numerical residual 
grows as the pole is approached. We colored the regions $R_1$ (yellow) $R_2$ (orange) and $R_3\cup R_4$ (red) of the approximate solution.
    } 
    \label{fig:Delta}
\end{figure}

Figure~\ref{fig:delta-residual} shows $|\Delta|$ computed using $X_0$ and $X_{\text{num}}$. 
At early times both representations satisfy the equation to high accuracy, with residuals below $10^{-10}$ as expected.  
However, as the solution approaches the pole, the residual associated with the numerical solution grows and reaches $\mathrm{O}(10^{-2})$ near $T=3$. This behavior reflects the intrinsic difficulty of numerically integrating a nonlinear equation whose solution develops a nearby singularity: truncation and round-off errors are amplified by the rapidly growing derivatives~\footnote{This deterioration is not fundamental. More sophisticated numerical approaches, such as puncture-type methods that explicitly account for the singular behavior, could likely improve the numerical accuracy in this regime.}.
In contrast, the residual computed from the analytic approximate solution remains uniformly bounded throughout the entire transition regime, staying below $10^{-7}$ up to $T_0-\epsilon$. 

This behavior is even more pronounced for the residual 
$\widetilde{\Delta}$, shown in Fig.~\ref{fig:tilde-delta-residual}.
While the residual associated with the numerical solution (black dashed line) remains below
$\mathrm{O}(10^{-3})$ in regions $R_1$ and $R_2$, it grows above $\mathrm{O}(1)$ in regions $R_3\cup R_4$. By contrast, the residual of the approximate solution (green line) remains uniformly bounded below 
$\mathrm{O}(10^{-5})$ across the entire transition. The degradation in the numerical residual arises from the combination of integrating the numerical solution (through the term $Y$) and differentiating it repeatedly, operations whose accuracy deteriorates as the singularity is approached, as discussed above.

The stability of the approximate solution is not accidental but follows directly from the controlled error bounds built into the construction of $X_0$. The analytic representation therefore provides uniform control over both the solution and its derivatives precisely in the regime where numerical integration becomes increasingly unreliable (without the use of more sophisticated numerical methods, as is the case in current implementations~\cite{Kuchler:2024esj,pc-Lorenzo}). We emphasize, however, that the numerical solution itself remains accurate for constructing the transition-to-plunge trajectories: the difference between the numerical and approximate solutions is at the level of $10^{-10}$. The loss of accuracy becomes significant only when repeated differentiation or integration is required near the singularity. 
This effect might then become relevant when computing subleading corrections to the transition dynamics, where derivatives of $X$ enter explicitly (see e.g., Eq.~(135) in Ref.~\cite{Compere:2021iwh}). A more in depth analysis of the improvement in the higher order corrections to the transition dynamics would need to be quantified. However, in principle, the bounded analytic error of $X_0$ provides a moderate advantage, and unless stated otherwise, we will take $X_0$ as the solution of the Painlev\'e I differential equation in this work.

\section{Transition to plunge trajectories}

The transition solution cannot be used to describe the trajectory all the way till the small body is absorbed by the larger black hole. This is evident from the fact that the radius becomes negative for $T\to T_0$ (see Fig.~\ref{fig:X(T)}), due to the pole-dominated behavior of the transition solution. As noticed in {Refs.~\cite{Buonanno:2000ef,Ori:2000zn}}, this behavior signals the breakdown of the applicability of the transition regime, and the onset of the geodesic plunge.

To describe the trajectory up to the moment when the compact object crosses the horizon, one must append a geodesic plunge for $0 \ll T \ll T_0$, corresponding to some point in region $R_2$ of Fig.~\ref{fig:X(T)}. One approach to constructing such a trajectory is \emph{patching}: the values of the energy, angular momentum, and Carter constant are frozen at a chosen time $T_{\text{patch}}$ within $R_2$, and the remainder of the trajectory for $T > T_{\text{patch}}$ is replaced by a plunging geodesic with the corresponding constants of motion $(E_{\text{patch}},\, L_{\text{patch}},\, Q_{\text{patch}})$. This approach has been employed in, e.g., Ref.~\cite{Apte:2019}. Although the resulting trajectory depends on the choice of $T_{\text{patch}}$, it is shown in Ref.~\cite{Lim:2019xrb} that this sensitivity does not affect the quasinormal modes excited during ringdown.

Nevertheless, to avoid this dependence on the patching time altogether, we instead adopt a \emph{matching} procedure to construct the transition-to-plunge trajectories as done in Refs.~\cite{Kuchler:2024esj,Compere:2021zfj} for a Schwarzschild background. The matching procedure for the radial motion is detailed below.  
We begin by constructing 
the radial motion from the ISSO, described by the transition solution, to the moment the compact object dives into the horizon, by matching the transition and geodesic plunge solutions in their common region of validity. The radial motion during the transition follows from Eqs.~\eqref{eq:expansion-variables-transition:1} and \eqref{eq:transformation-X_T}, and is given by
\begin{align}
    r_{\text{trans}}(\lambda) = &r_{\text{ISSO}}  \\
    &+\eta^{2/5} A^{-3/5} B^{2/5}  X_0[-(\eta A B)^{1/5}(\lambda-\lambda_{\text{ISSO}})]\,, \notag
\end{align} with $X_0$ explicitly given in Eq.~\eqref{eq:approx-exact-sol-X0}. 
This expression is valid throughout the transition regime, where the rescaled time $T=-(\eta AB)^{1/5}(\lambda-\lambda_{\text{ISSO}})$ remains $\mathrm{O}(1)$.

The radial motion during the geodesic plunge can be obtained from Eq.~\eqref{eq:plunge-eq}. Although $r=r_{\text{ISSO}}$ is a solution to that equation, we are interested in the nontrivial branch asymptoting to the ISSO from below, $r<r_{\text{ISSO}}$. To isolate this solution, we introduce
\begin{equation}\label{eq:transformation-plunge}
    r(\lambda) = r_{\text{ISSO}}-x_{\text{plunge}}(\lambda)\,,
\end{equation} where the minus sign with respect to Eq.~\eqref{eq:expansion-variables-transition:1} ensures that $0\leq x_{\text{plunge}}(\lambda)\leq r_{\text{ISSO}}-r_4$.~\footnote{The upper limit in $x_{\rm plunge}$ has been chosen so that the right hand side of Eq.~\eqref{eq:plunge-eq} is positive.} Note $x_{\rm plunge}$ should not be confused with $x$.
 Substituting into Eq.~\eqref{eq:plunge-eq} yields
\begin{equation}\label{eq:x-plunge-diff-eq}
    \left(\frac{\d x_{\text{plunge}}}{\d\lambda}\right)^2 = (1-E_{\text{ISSO}}^2) x_{\text{plunge}}^3(r_{\text{ISSO}}-r_4-x_{\text{plunge}})\,,
\end{equation} which can be integrated exactly. The resulting solution is
\begin{equation}\label{eq:x-plunge}
    x_{\text{plunge}} (\lambda) = \frac{4(r_{\text{ISSO}}-r_4)}{4+(1-E_{\text{ISSO}}^2) (r_{\text{ISSO}}-r_4)^2(\lambda-k)^2}
\end{equation} where $k$ is an integration constant. The radial motion during the plunge is therefore given by Eq.~\eqref{eq:transformation-plunge} combined with Eq.~\eqref{eq:x-plunge}. This solution remains valid as long as $r_{\text{ISSO}}-r_4-x_{\text{plunge}}\geq0$.

The matching between the transition and plunge solutions must be performed in the region where both descriptions apply. The transition solution is valid for rescaled times $T=\mathrm{O}(1)$, corresponding to radial deviations $r-r_{\text{ISSO}}\sim \eta^{2/5}$. The plunge solution, on the other hand, is the exact geodesic solution obtained by freezing the constants of motion at their ISSO values. To match the two descriptions, we consider the early-time behavior of the plunge solution, corresponding to $x_{\text{plunge}}\to0^+$ (i.e., $r\to r_{\text{ISSO}}^-$), and the late-time behavior of the transition solution $T\to T_0^-$. These limits overlap in the regime
\begin{equation}\label{eq:validity}
    0\gg \lambda-\lambda_{\text{ISSO}} \gg -(\eta A B)^{-1/5} T_0\,.
\end{equation}  

To carry out the matching, we therefore examine the asymptotic behavior of both solutions in this overlap region. For the transition solution, taking the limit $T\to T_0^-$ while remaining in the overlapping regime of validity in Eq.~\eqref{eq:validity} places us in region $R_2$ of the approximate Painlevé solution. From Eq.~\eqref{eq:approx-exact-sol-X0}, the leading behavior is
\begin{equation}
    X_0(T) \sim-\frac{6}{(T-T_0)^2}\,.
\end{equation}Substituting this into $r_{\text{trans}}$ and using Eq.~\eqref{eq:transformation-X_T}, we obtain
\begin{equation}\label{eq:trans-right}
    \underrightarrow{r_{\text{trans}}}(\lambda) = r_{\text{ISSO}} -\frac{6}{A \left(\lambda-\lambda_{\text{ISSO}} +\frac{T_0}{(\eta A B)^{1/5}}\right)^2}
\end{equation} where $\underrightarrow{\quad}$ denotes the asymptotic behavior of this solution as $T\to T_0^-$, and $A$ was given in Eq.~\eqref{eq:A-B-def}.

Next, we consider the plunge solution in the limit $r\to r_{\text{ISSO}}^-$, or equivalently $x_{\text{plunge}}\to 0^+$. Expanding Eq.~\eqref{eq:x-plunge-diff-eq} for small $x_{\text{plunge}}$ and integrating yields
\begin{align}
    -\frac{2}{\sqrt{r_{\text{ISSO}} -r_4}\sqrt{x_\text{plunge}}} + \mathrm{O}(x_{\text{plunge}}^{1/2}) = \notag
    \\
    \pm\sqrt{1-E^2_{\text{ISSO}}} (\lambda-k)\,.
\end{align}  
Solving for $x_{\text{plunge}}$ and substituting into Eq.~\eqref{eq:transformation-plunge} gives
\begin{equation}\label{eq:plunge-left}
    \underleftarrow{r_{\text{plunge}}} (\lambda) = r_{\text{ISSO}} -\frac{4}{(1-E_{\text{ISSO}}^2) (r_{\text{ISSO}}-r_4)(\lambda-k)^2}\,,
\end{equation} where the under-arrow indicates that this is an asymptotic solution, and $k$ is an integration constant.

For the matching to be possible, $\underrightarrow{r_{\text{trans}}}$ and $\underleftarrow{r_{\text{plunge}}}$ must share the same functional dependence. Using Eqs.~\eqref{eq:A-B-def} and~\eqref{eq:plunge-eq}, one finds that the constant $A$ is 
\begin{equation}
    A = \frac{3}{2} (1-E_{\text{ISSO}}^2) (r_{\text{ISSO}}-r_4)\,.
\end{equation}which ensures that Eq.~\eqref{eq:trans-right} has the same constants in front of the inverse–square structure as Eq.~\eqref{eq:plunge-left}. In fact, the two expressions coincide upon identifying 
\begin{equation}
    k=\lambda_{\text{ISSO}} {- }\frac{T_0}{(\eta A B)^{1/5}}\,.
\end{equation}

A uniformly valid approximation across the transition and plunge regimes is then obtained through the standard composite construction
\begin{equation}
    r_{\text{matched}} (\lambda) = r_{\text{trans}}(\lambda) + r_{\text{plunge}}(\lambda) -\underrightarrow{r_{\text{trans}}}(\lambda)\,.
\end{equation}

\begin{figure}
    \centering
    \includegraphics[width=\linewidth]{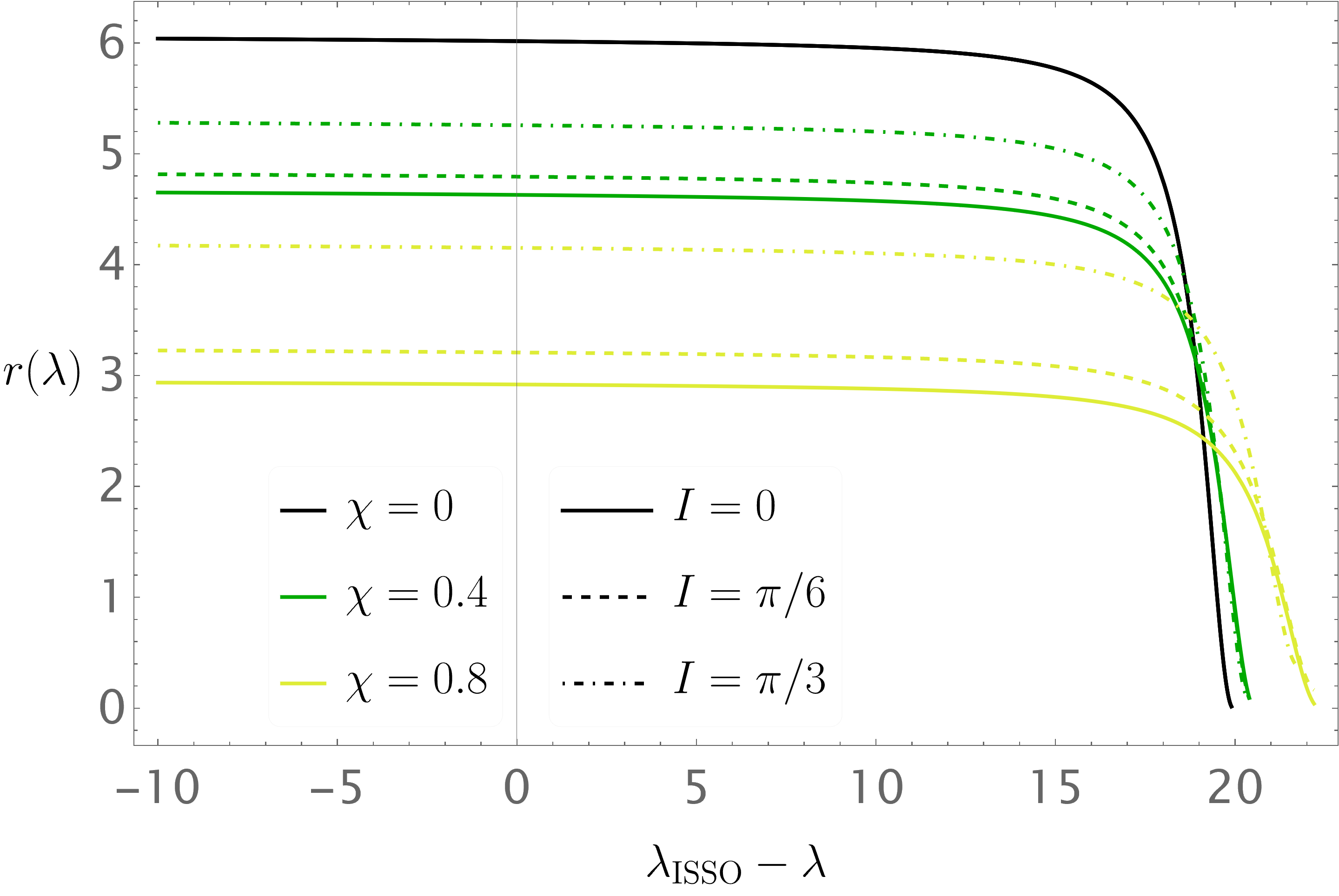}
    \caption{Composite transition and geodesic plunge solutions as a function of Mino time. We show the trajectories for three values of the dimensionless spin of the central black hole $\chi=0\,,0.4\,,0.8$ and three inclinations $I=0\,,\pi/6\,,\pi/3$ for prograde orbits. The mass ratio has been fixed to $\eta = 10^{-5}$.  }
    \label{fig:trajectories}
\end{figure}

In Fig.~\ref{fig:trajectories} we present the composite matched solutions describing the transition and geodesic plunge for radiation-reaction-driven inspirals at leading order. These trajectories are not expected to remain valid at sufficiently early times, where the evolution is instead described by the adiabatic inspiral regime. Constructing the full global solution would therefore require matching the transition solution to an adiabatic inspiral at early times, following the procedure described in Refs.~\cite{Compere:2021zfj,Kuchler:2024esj}. Here, however, we restrict attention to the matched transition and plunge solutions only.

The trajectories are shown for different values of the dimensionless spin of the central black hole, $\chi=a/M$, and orbital inclinations $I$, while fixing the mass ratio to $\eta=10^{-5}$. Each color corresponds to a different spin ($\chi=0$, $0.4$, and $0.8$ shown in black, green, and yellow, respectively), while the inclination is indicated by the line style: solid for $I=0$, dashed for $I=\pi/6$, and dot-dashed for $I=\pi/3$.
The sharpest transition occurs for the nonrotating case, for which all inclinations yield the same trajectory, as expected from the spherical symmetry of the Schwarzschild spacetime. For spinning black holes this degeneracy is broken, and larger inclinations shift the transition to larger radii. Increasing the spin also broadens the transition regime, leading to a more gradual trajectory, in which the compact object takes longer to dive into the horizon. 

To generate Fig.~\ref{fig:trajectories} we computed the coefficients $A$ and $B$ defined in Eq.~\eqref{eq:A-B-def}. This requires the rates of change of the energy, angular momentum, and Carter constant at the ISSO, namely the coefficients $\kappa_\mathcal{C}$ introduced in Eq.~\eqref{eq:C-evolution}. The rates of change of the energy and angular momentum with respect to Boyer--Lindquist coordinate time were provided by Maarten van de Meent in the form of \texttt{Mathematica} interpolating functions. The conversion to Mino time was performed using the \texttt{KerrGeoFrequencies} routine from the Black Hole Perturbation Toolkit~\cite{BHPToolkit}. The evolution of the Carter constant was obtained from the quasi-circularity condition in Eq.~\eqref{eq:second-order-transition:2}. In Table~\ref{tab:kappas-A-B} we summarize the numerical values of the coefficients $\kappa_\mathcal{C}$ together with the corresponding values of $A$ and $B$ for different spins and inclinations.

   \begin{table}[]
\begin{tabular}{c|c|c|c|c|c|c|}
\cline{2-7}
                                        & $\chi$ & $\kappa_E$ & $\kappa_L$ & $\kappa_Q$  & $A$ & $B$   \\ \hline
\multicolumn{1}{|l|}{} & 0& 0.048& 0.704& $1.6\cdot10^{-7}$& 1& {}14.624  \\ \cline{2-7} 
\multicolumn{1}{|l|}{$I=0$}                  &0.4& 0.098&1.006& $3.2\cdot10^{-9}$& 1& {}12.573 \\ \cline{2-7} 
\multicolumn{1}{|l|}{}                  & 0.8& 0.312& 1.793& $-1.9\cdot 10^{-8}$& 1& {}8.250 \\ \hline
\multicolumn{1}{|l|}{} & 0& 0.048& 0.656& 0.936& 1& {}14.624 \\ \cline{2-7} 
\multicolumn{1}{|l|}{$I=\pi/6$}                  &0.4& 0.088&0.895& 1.032& 0.993& {}12.785  \\ \cline{2-7} 
\multicolumn{1}{|l|}{}                  &0.8 & 0.235 & 1.474 & 1.065 &0.936 & {}9.017 \\ \hline
\multicolumn{1}{|l|}{} & 0& 0.048& 0.528& 3.047& 1& {}14.624 \\ \cline{2-7} 
\multicolumn{1}{|l|}{$I=\pi/3$}                  &0.4& 0.066&0.638& 3.097& 0.982& {}13.303 \\ \cline{2-7} 
\multicolumn{1}{|l|}{}                  &0.8 & 0.110 & 0.844 & 2.983 & 0.882 & {}10.691  \\ \hline
\end{tabular}
\caption{Rates of change of the energy, angular momentum, and Carter constant at the ISSO for three values of the spin of the central black hole and three inclinations. We restrict ourselves to prograde orbits. The rates change of the energy, angular momentum, and Carter constant as a function of Boyer-Lindquist time at the ISSO were provided by Maarten van de Meent. }
\label{tab:kappas-A-B}
\end{table}

\section{Universality of the transition to plunge}
\label{sec:universality}

 Up to this point, our focus has  been on the solutions of the transition equation. However, the derivation of the transition equation itself, reviewed in Sec.~\ref{subseec:transition-to-plunge}, relied on a number of physical assumptions. 
In particular, in Eq.~\eqref{eq:second-order-transition:1} it is assumed that the dissipative component of the self-force enters only through a slow evolution of the geodesic constants of motion, that the conservative piece merely shifts the ISSO location (as discussed in Ref.~\cite{Ori:2000zn}) and can therefore be neglected at leading order, and that the dominant evolution of the constants of motion is linear in time, cf. Eq.~\eqref{eq:C-evolution} (see also Refs.~\cite{Ori:2000zn,Apte:2019}). While these assumptions are physically well motivated, they remain approximations. One may therefore ask whether modifying the precise time dependence of the evolving constants of motion or relaxing the assumption that the self-force acts exclusively through slow parameter drift, would alter the differential equation governing the transition.

In Sec.~\ref{subsec:generalization}, we address this question by examining the robustness of the transition equation under admissible perturbations. We introduce controlled modifications of the differential equation and apply the Painlevé test to determine whether the perturbed equations retain the Painlevé property, namely that their movable singularities are poles. The persistence of this property indicates that the transition dynamics remains within the Painlevé I universality class.

Beyond the robustness of the equation itself, we emphasize that the emergence of Painlevé I persists even when the orbital configuration is generalized from equatorial to inclined Kerr geodesics. Despite the increased complexity of the parameter space, the transition dynamics reduces locally to the same equation. This strongly suggests that the appearance of Painlevé I reflects a universal mechanism describing the transition to plunge.

To clarify the geometric origin of this universality, we turn in Sec.~\ref{subsec:catastrophe-theory} to catastrophe theory. Catastrophe theory{, introduced in the seminal work of Ren\'e Thom in 1972~\cite{Thom1972},} classifies the structurally stable degeneracies of critical points of smooth functions under variation of control parameters. In the present context, the effective radial potential defines a family of equilibria parameterized by the orbital constants and the black-hole spin. The ISCO corresponds to a degenerate critical point at which a stable and an unstable circular orbit merge and form an inflection point. Such a merger is the hallmark of a fold catastrophe. The transition to plunge may therefore be interpreted as the slow evolution of the system across the fold curve of the catastrophe manifold.

\subsection{Structural stability of the transition equation }
\label{subsec:generalization}
In this section we investigate whether analytic modifications of the transition equation preserve the Painlevé property. Such corrections may arise from higher-order terms in the time evolution of the constants of motion, from additional contributions to the self-force that might not have been taken into account in the derivation of the transition, or from small environmental effects that slightly perturb the dynamics near the ISSO. Our goal is to determine whether the characteristic singularity structure of the Painlevé I equation is robust under a controlled class of perturbations.

We consider the modified equation
\begin{equation}\label{eq:Painleve-modified}
 E\equiv    \frac{\d^2 x}{\d\lambda^2} +A x^2 +B\eta F(\lambda) =0
\end{equation} where
\begin{equation}\label{eq:F}
    F(\lambda) = \sum_{i=1}^\infty c_i (\lambda-\lambda_{\text{ISSO}})^i\,, \quad c_1=1\,,
\end{equation}
and the series is assumed to be absolutely convergent in a neighborhood of the ISSO. When $c_i=0$ for all $i\geq2$, Eq.~\eqref{eq:Painleve-modified} reduces to the transition equation~\eqref{eq:Painleve-eta}, which can be mapped to the Painlev\'e~I differential equation~\eqref{eq:PainleveI}, as discussed in Sec.~\ref{subseec:transition-to-plunge}. The class of modifications considered here is sufficiently general to incorporate higher-order terms in the Taylor expansion governing the slow evolution of the constants of motion, as well as more general analytic corrections to the forcing term proportional to $\eta$.  A subset of these corrections was previously considered in Ref.~\cite{Apte:2019}, where polynomial corrections up to cubic order in Mino time were introduced to model the evolution of the constants of motion during the transition.

To assess whether this modified equation retains the Painlevé property, we apply the Painlevé test in the Kowalevski–Gambier formulation~\cite{Kowalevski1889,Painleve-handbook}. The test examines whether the general solution can be expressed locally as a Laurent expansion about a movable singularity with only pole-type behavior. Passing the test is a necessary condition for the Painlevé property and provides strong evidence, though not a proof, of integrability in the Painlevé sense. We follow the exposition of the test described in Ref.~\cite{Painleve-handbook}. 
We start by assuming the existence of a movable singularity at $\lambda=\lambda_0$ and introduce $\chi=\lambda-\lambda_0$ (not to be confused with the dimensionless spin variable). We look for the dominant singular behavior of the form 
\begin{equation}
    x\underset{\chi\to 0}{\sim} x_0\chi^p\,,\quad x_0\neq 0\,, \quad -p\in \mathbb{N}\,.
\end{equation}
Substituting into the differential equation yields the leading scalings
\begin{equation}
    x^{\prime\prime} \sim p(p-1) x_0 \chi^{p-2}\,,\quad x^2\sim x_0^2 \chi^{2p}\,.
\end{equation} Since $F(\lambda)$ is analytic in $\lambda$, it is also analytic in $\chi$, and therefore contributes only non-negative powers of $\chi$. Consequently it cannot participate in the dominant singular balance. The only possible balance between negative powers is obtained by equating the exponents of $x^{\prime\prime}$ and $x^2$, which gives 
\begin{equation}
    p-2=2p\Rightarrow p=-2\,.
\end{equation} A necessary condition for the Painlev\'e property is that the leading power $p$ is a negative integer, which in this case is satisfied. Thus any movable pole-type singularity must be a double pole, in agreement with the behavior of solutions to Painlevé I.

Matching the coefficients at leading order yields
\begin{equation}\label{eq:x0}
    p(p-1) x_0  +A x_0^2 = 0\Rightarrow x_0 = -\frac{6}{A}\,.
\end{equation} The dominant singular structure is unchanged by the analytic forcing term $F(\lambda)$.

After the completion of this first step, one must determine whether a Laurent series of the form 
\begin{equation}\label{eq:LaurentSeries}
    x=\sum_{j=0}^\infty x_j \chi^{j+p}\,,\quad E=\sum_{j=0}^\infty E_j \chi^{j+q}\,,\quad -p,-q\in\mathrm{N}\,.
\end{equation} exists and whether all the coefficients in the series can be computed recursively. Hence, in the {second step} of the test we compute the indicial equation and Fuchs numbers of the linearized ODE, which indicates at which order in the expansion of the solution arbitrary constants can enter. The determination of such constants imposes constraints on the coefficients of the differential equation, thus allowing us to answer the question of whether Eq.~\eqref{eq:Painleve-modified} passes the Painlev\'e test for arbitrary coefficients $c_i\,, i\geq 2$.  The dominant terms of the differential equation, denoted by $\widehat{E}[x]$, are
\begin{equation}
\widehat{E}[x] =    x^{\prime\prime} + A x^2\,. 
\end{equation} The linearization of the dominant singular terms around the leading order solution is obtained by adding the next-to-leading order correction, i.e.,  $x\sim x_0 \chi^{-2} + \epsilon X$, and yields
\begin{equation}
    \lim_{\epsilon\to0} \frac{1}{\epsilon}(\widehat{E}[x+\epsilon X]-\widehat{E}[x]) =X^{\prime\prime} +2A x X = 0\,.
\end{equation} Looking for power-law solutions of the form $X\sim x_j \chi^{j-2}$ for $j$ a positive integer,  leads to the indicial equation 
\begin{equation}
      x_j \chi^{j-4} (j+1)(j-6)=0\,.
\end{equation}  The associated Fuchs indices are therefore $j=-1$ and $j=6$. The index $j=-1$ corresponds to the arbitrariness of the pole position $\lambda_0$, while $j=6$ indicates that a second arbitrary constant may enter at order $x_6\chi^{4}$.

We now construct the full Laurent expansion around the movable singularity
\begin{equation}
    x=\sum_{j=0}^\infty x_j \chi^{j-2}\,,\quad E=\sum_{j=0}^\infty E_j \chi^{j-4}\,,
\end{equation} and substitute it into the complete modified equation~\eqref{eq:Painleve-modified}. We find the coefficients $x_j$ order by order up to $j=6$ to check whether  obstructions exist to the construction of the power series solution. 

Matching coefficients order by order in $\chi$ shows that the recursion uniquely determines $x_j$ for $j=1\,,...\,5$. In particular, we find that for $x_0$ the recurrence relation is automatically satisfied once we account for Eq.~\eqref{eq:x0}, while for $j=1\,,...\,,5$
\begin{subequations}\label{eq:recurrance-rel}
    \begin{align}\label{eq:recurrance-rel:j}
          (E_j=0): &\quad x_j =0 \text{ for } j=1\,,2\,,3 \\\label{eq:recurrance-rel:4}
          (E_4=0): &\quad x_4 =\frac{B\eta}{10} \sum_{i=1}^\infty c_i(\lambda_0-\lambda_{\text{ISSO}})^i \\\label{eq:recurrance-rel:5}
          (E_5=0): &\quad x_5 =\frac{B\eta}{6} \sum_{i=1}^\infty c_i i (\lambda_0-\lambda_{\text{ISSO}})^{i-1} 
    \end{align}
\end{subequations}

At $j=6$, however, the coefficient multiplying $x_6$ vanish identically. Instead of determining $x_6$, one then obtains a compatibility condition involving the forcing function $F(\lambda)$, namely
\begin{equation}
    \label{eq:recurrance-rel:6}
          (E_6=0): \quad \sum_{i=1}^\infty c_i\frac{i(i-1)}{2} (\lambda_0-\lambda_{\text{ISSO}})^{i-2} =0\,.
\end{equation}
If this condition is satisfied, the recursion proceeds consistently, $x_6$ remains arbitrary, and higher-order coefficients can be determined recursively without further obstructions. The local solution then depends on the two arbitrary constants $(\lambda_0\,, x_6)$, and no logarithmic terms appear in the expansion. If the condition fails, logarithmic branching is generically expected, and the equation does not pass the Painlev\'e test.   

Notice that for $c_j=0$ for $j\geq 2$ the constraint is automatically satisfied, and hence Eq.~\eqref{eq:Painleve-eta} passes the Painlev\'e test, as expected. Furthermore, affine forcing terms of the form $c_0+c_1 (\lambda-\lambda_{\text{ISCO}})$ automatically pass the test as well~\footnote{Although we set $c_0 = 0$ in Eq.~\eqref{eq:F}, allowing $c_0\neq0$ similarly leads to Eq.~\eqref{eq:recurrance-rel}, with the series coefficients $x_i$ acquiring additional terms proportional $c_0$. However, the outcome of the test remains unchanged.   }.

For $c_j\neq 0$ for $j\geq 2$ the forcing term $F(\lambda)$ must satisfy Eq.~\eqref{eq:recurrance-rel:6}. Under the classical interpretation of the Painlevé test, this condition signals a failure of the test: the constraint on the coefficients depends on $\lambda_0$ -- the location of the movable singularity -- and therefore on the initial conditions. In other words, this means that if the constraint is satisfied, the coefficients in the original differential equation depend on the initial conditions. This is indirect contradiction with the Painlev\'e property.
However, if we restrict ourselves to the subclass of solutions corresponding to the tritronquée solutions (which recall are the physically relevant solutions), then this fixes the singularity location and the Painlev\'e test is satisfied for this subclass. Thus, if the constraint in Eq.~\eqref{eq:recurrance-rel:6} is satisfied, the Painlev\'e test is satisfied in this weaker sense. Consequently, for this distinguished class one may still construct a Laurent expansion of the perturbed equation~\eqref{eq:Painleve-modified} with a general forcing term satisfying Eq.~\eqref{eq:recurrance-rel:6}.

For physical perturbations, however, the relevant transition equation need not be Eq.~\eqref{eq:Painleve-modified}: because the transition is governed by a slow-time, the leading dynamics is organized in terms of the rescaled variable $T\sim \eta^{1/5}(\lambda-\lambda_{\text{ISSO}})$ rather than $\lambda$. In that setting, condition~\eqref{eq:recurrance-rel:6} (i.e., the resonance compatibility requirement) is not expected to control the physical behavior. Instead, any higher-order terms in the analytic forcing $F(\lambda)$ enter as subleading corrections in the small$-\eta$ expansion, appearing as higher-order perturbations of the Painlev\'e~I equation, in the same spirit as Ref.~\cite{Compere:2021zfj}. Thus, for many physical situations, the leading singular behavior governing the transition remains unchanged, underscoring the rigidity of the transition.

The present analysis applies only to analytic forcing terms depending on $\lambda$ alone. More general perturbations involving nonlinear powers of $x$ beyond quadratic order or derivative-dependent terms may alter the dominant balance and thereby change the singularity structure of the equation. Establishing the Painlevé property in such cases would require a separate analysis. Terms of the form $x^2$ and $x \lambda$ could be added without changing the singular behavior of Eq.~\eqref{eq:Painleve-modified}, although the coefficients in the Laurent series expansion would change. 
 Within the analytic class considered here, however, we have shown explicitly that the modified transition equation preserves the pole-type movable singularities and resonance structure of Painlevé I, provided the resonance compatibility condition in Eq.~\eqref{eq:recurrance-rel:6} is satisfied.

\subsection{A catastrophic interpretation of the transition to plunge}
\label{subsec:catastrophe-theory}

Catastrophe theory{~\cite{Thom1972}} studies the structurally stable critical points of smooth systems and the universal ways in which these equilibria change under smooth variation of parameters. Remarkably, the local behavior around these singular points, the catastrophe singularities, is highly constrained and falls into a small number of universal classes. Interpreting the transition-to-plunge as the slow passage through one such catastrophe explains two key features of the dynamics: first, why the transition equation is structurally robust under perturbations, and second, why the transition for inclined Kerr orbits remains qualitatively as simple as the equatorial Schwarzschild case. In both situations, the dynamics is governed by the same underlying catastrophe structure.

In the following, we first review the basic concepts of catastrophe theory in Sec.~\ref{subsubsec:catastrophe-theory-intro}. We then show in Sec.~\ref{subsec:Kerr-parameter-space} that the space of structurally stable equilibria for equatorial Kerr orbits can be mapped to the fold catastrophe, while the equilibrium manifold for inclined Kerr orbits is described by the cusp catastrophe. Finally, in Sec.~\ref{sec:transition-fold}, we show that the transition equation reduces to the Painlev\'e~I differential equation because, in both cases, the transition corresponds to the slow crossing of a fold catastrophe.
\subsubsection{A catastrophe theory primer}
\label{subsubsec:catastrophe-theory-intro}

Catastrophe theory provides a geometric framework for describing how equilibria of a smooth function change under variation of external parameters~\cite{zeeman1977catastrophe}. { In this subsection we follow} { partially the presentation in~\cite{Jaramillo:2022mkh}, in particular its notation.} Let $\Phi(\mathbf{R}\,;\mathbf{X})$ 
be a smooth function depending on a set of state variables $\mathbf{R}\in S$ and control parameters $\mathbf{X}\in D$. The state variable space $S$ is a assumed to be $m$-dimensional,  while the control parameter space $D$ is $n$-dimensional. These two spaces can a priori be infinite, but in practice, only the relevant parameters for the caustic structure need to be accounted for~\cite{Jaramillo:2022mkh}. In fact, for most physical situations, a finite number of state variables and control parameters, with  $n\geq m$ suffices.

The smooth function $\Phi:S\times D\to\mathbb{R}$ is called the generating function, and in the present context, it plays the role of an effective potential~\footnote{The generating function is a more general concept than the effective potential. In gravitational lensing, for example, the time delay plays the role of a generating function~\cite{Jaramillo:2022mkh}. More broadly, catastrophe theory has found applications beyond physics, including in the social sciences, with models in psychology and finance~\cite{zeeman1977catastrophe}.}: equilibria are obtained from
\begin{equation}\label{eq:equilibria-eq-catastrophe}
    \nabla_{\mathbf{R}} \Phi(\mathbf{R}\,;\mathbf{X})  =0\,, 
\end{equation} where the gradient is taken with respect to the state variables.

For fixed $\mathbf{X}$, solutions of this equation define the stationary points of the system. As $\mathbf{X}$ varies, these stationary points trace out an $n$-dimensional surface $M\subset S\times D$, known as the stationary (or catastrophe) manifold. The projection of $M$ onto control-parameter space $D$ need not be injective as a point in $D$ can have several preimages: several equilibria may correspond to the same set of control parameters. The projection map $\xi:M\to D$
is known as the catastrophe map. 

A catastrophe occurs when two or more stationary points coalesce and change their nature. Mathematically, this corresponds to a degenerate critical point of $\Phi$, where in addition to Eq.~\eqref{eq:equilibria-eq-catastrophe}, the Hessian with respect to the state variables becomes non-invertible~\footnote{In optics, the magnification matrix is usually defined as $\mathcal{M}_{\alpha\beta}^{-1}(\mathrm{X}) = -\left(\frac{\partial^2 \Phi}{\partial R_\alpha\partial R_\beta}\right)_{(X_1\,,...\,,X_n)}$.  The light intensity then scales as $I=|\det \mathcal{M}_{\alpha\beta}|$, so crossing a caustic leads to a divergence of the intensity~\cite{Jaramillo:2022mkh}. In the transition-to-plunge problem, however, there is no direct analogue of the light intensity, and therefore no corresponding divergent observable is expected at the catastrophe.}, i.e., 
\begin{equation}\label{eq:hessian}
    \det \left(\frac{\partial^2 \Phi}{\partial R_\alpha \partial R_\beta }\right) =0\,.
\end{equation}
The projection of such degenerate points into control-parameter space through the catastrophe map $\xi$ defines the caustic (or bifurcation set).

A central result due to Thom { and further developed by Arnol'd}~\cite{Thom1972,arnol1972normal} is that, up to smooth changes of variables, only a finite number of local normal forms describe all structurally stable degeneracies when the number of control parameters does not exceed four. For one state variable, the relevant catastrophes are the fold and the cusp. Their normal forms are
\begin{equation}\label{eq:fold-std}
    \phi_{\text{fold}}(\mathrm{s}\,;x_1) = \frac{\mathrm{s}^3}{3}-x_1 \mathrm{s}
\end{equation}
\begin{equation}\label{eq:cusp-std}
    \phi_{\text{cusp}}(\mathrm{s}\,;x_1\,,x_2) = \frac{\mathrm{s}^4}{4}+\frac{x_2}{2} \mathrm{s}^2 + x_1 \mathrm{s}
\end{equation} 
where $\mathrm{s}$ represents the state variable, and $x_1\,,x_2$ the control parameters. 
 The normal form of the generating function assumes that the highest order singularity lies at $\bm{x}=0$, and it can be obtained from any generating function $\Phi(\bm{R}\,,\bm{X})$ through appropriate diffeomorphisms of the state variables $\bm{R}\mapsto \bm{\mathrm{s}}(\bm{R})$ and control parameters $\bm{X}\mapsto \bm{x}(\bm{X})$. 
 The standard form of the catastrophe's generating function $\phi(\bm{\mathrm{s}}\,,\bm{x})$ is related to $\Phi$ by 
\begin{equation}\label{eq:standard form}
    \Phi(\bm{R}\,,\bm{X}) = C(\bm{X}) + \phi(\bm{\mathrm{s}}(\bm{R})\,,\bm{x}(\bm{X}))\,,
\end{equation} where $C(\bm{X})$ is a smooth function of the control parameters only, so the stationary points of $\Phi$ and $\phi$ coincide~\cite{Jaramillo:2022mkh}. 

\begin{figure}
    \centering
    \includegraphics[width=\linewidth]{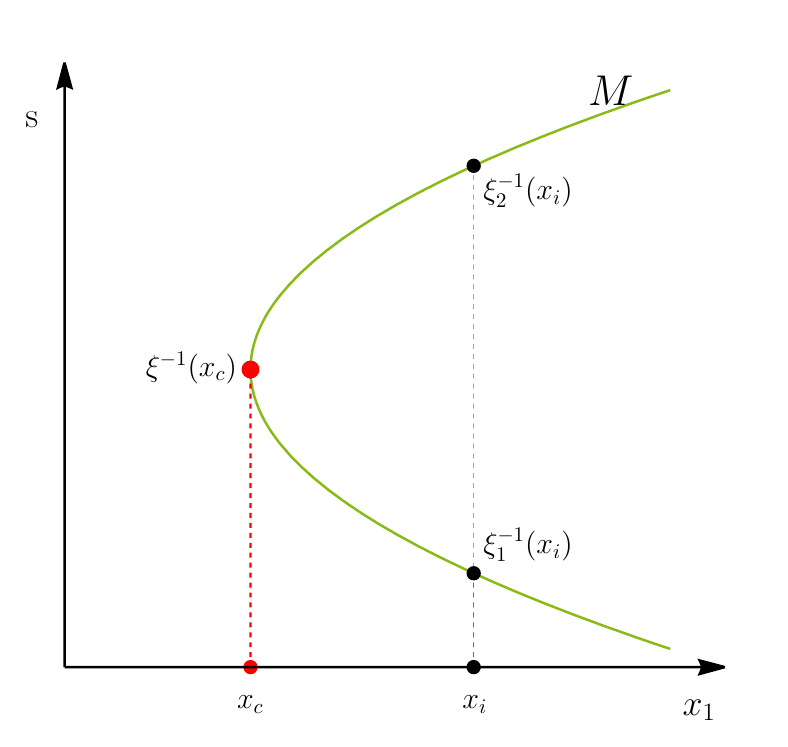}
    \caption{Fold catastrophe manifold $M\subset S\times D$. The caustic $x_c$ has been represented in the control parameter space $D$, which is one-dimensional and parametrized by $x_1$. The fold catastrophe is marked in red. The two preimages by the catastrophe map of a point $x_i\in D$ are shown in the catastrophe manifold by black dots.  }
    \label{fig:fold-general}
\end{figure}

The fold catastrophe is obtained when there is only one relevant control parameter. The cusp arises when the control parameter space is two-dimensional. 
In the fold catastrophe, two stationary points merge and annihilate as the control parameter crosses the caustic. The simplest representation of this phenomena is given by a parabola. This can be seen by solving Eq.~\eqref{eq:equilibria-eq-catastrophe} for the fold generating function in its standard form~\eqref{eq:fold-std}, which yields
\begin{equation}
    \nabla_{\mathrm{s}}\phi_{\text{fold}} = \mathrm{s}^2-x_1=0 \quad \Rightarrow \quad  \mathrm{s}=\pm \sqrt{x_1}\,.
\end{equation} For $x_1>0$ there are two distinct equilibria, while for $x_1=0$ they merge. For $x_1<0$ no critical points exist. In Fig.~\ref{fig:fold-general} we show the fold in its standard form and mark the relevant points for catastrophe theory.   

In the cusp catastrophe, the fold curve itself acquires a singular point, organizing a two-parameter family of folds. The catastrophe manifold is obtained by solving Eq.~\eqref{eq:equilibria-eq-catastrophe} with the standard form for the cusp generating function. This yields
\begin{equation}\label{eq:equilibrium-cusp}
    \nabla_\mathrm{s} \phi_{\text{cusp}} =\mathrm{s}^3+x_2 \mathrm{s}+x_1=0\,,
\end{equation} which is a depressed cubic polynomial. The roots of the polynomial become degenerate when the discriminant
\begin{equation}
    -(4 x_2^3+27 x_1^2)=0
\end{equation} vanishes. This occurs along the lines $x_1=\pm \frac{2}{3}\sqrt{-\frac{x_2^3}{3}}$, for which two of the roots of the cubic become degenerate, and hence for which the Hessian defined in Eq.~\eqref{eq:hessian} vanishes. These are fold lines~\footnote{A generic transverse $(n-1)$-dimensional slicing of an $n$-dimensional catastrophe manifold produces a $(n-1)$-dimensional catastrophe manifold. In other words, locally, a generic 1-dimensional slice of a cusp is a fold~\cite{Thom1972,zeeman1977catastrophe}. }.

\begin{figure}
    \centering
   \includegraphics[width=\linewidth]{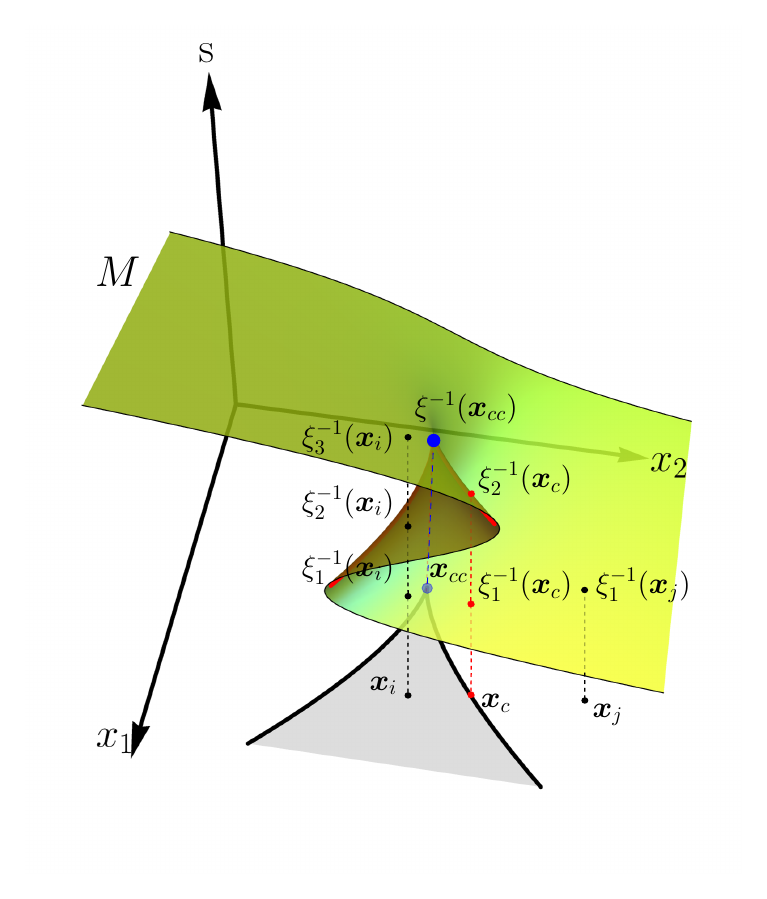}
        \caption{Cusp catastrophe manifold embedded in $S\times D$. We show four points in $D$, $\bm{x}_i$ within the cusp region, $\bm{x}_j$ outside of the cusp region, $\bm{x}_c$ on the fold caustic (red), and $\bm{x}_{cc}$ in cusp caustic (blue). Their respective preimages on the manifold $M$ have been indicated.  The cusp catastrophe, where the three sheets join on $M$, has also been marked in blue. }
    \label{fig:cusp-general}
\end{figure}

Furthermore, the roots of the polynomial in Eq.~\eqref{eq:equilibrium-cusp} become triply degenerate at the point $\bm{x}=0$. This is where the cusp catastrophe occurs. The projections of these points to control parameter space are the fold and the cusp caustics. The shaded region in control parameter space in Fig.~\ref{fig:cusp-general} marks the region of parameters where the generating function has three critical points, and the white segment the region where only one such equilibria exists. 

In what follows, we will show that both the ISCO and the ISSO corresponds to such a fold degeneracy of the Kerr effective potential, and that the transition to plunge can be interpreted as slow evolution across the associated caustic.

\subsubsection{The Kerr parameter space as a cusp manifold}
\label{subsec:Kerr-parameter-space}

We now use the tools of catastrophe theory summarized in the previous section to analyze the space of equilibrium solutions of Kerr geodesics. While the structure of Kerr geodesic phase space has been extensively studied in the literature (see e.g.~\cite{Chandrasekhar:1985kt,Warburton:2013,Compere:2022-Kerr-geo,Stein_2020}), 
here we emphasize its interpretation in terms of universal catastrophe morphologies.

To study the space of equilibrium configurations we take as generating function the radial effective potential,
\begin{equation}\label{eq:generating-function-Kerr}
    \Phi(\mathbf{R}\,;\mathbf{X}) = V_{\text{eff}} (r\,; E\,,L\,,Q)\,.
\end{equation}
This choice is natural because equilibrium states of the radial motion correspond to critical points of $V_{\text{eff}}$. Catastrophe theory classifies the degeneracies of such critical points under variation of control parameters.

Although in principle the full dynamics involves both $r$ and $\theta$, the polar potential does not develop higher-order degeneracies in the regime relevant for the transition to plunge. The singular structure is therefore entirely governed by the radial potential and considering the $\theta$ motion would not change the morphology of the catastrophe manifold.

The effective potential for the Kerr radial geodesic motion can be written as
\begin{equation}\label{eq:def-Veff}
    V_{\text{eff}}(r) \equiv E^2-\frac{R(r)}{r^4}\,,
\end{equation} which explicitly reads
\begin{equation}\label{eq:def-Veff-explicit}
\begin{split}
    V_{\text{eff}}(r) &= 1-\frac{2M}{r}
   +\frac{L^2+Q+a^2(1-E^2)}{r^2}\\
   &-2M\frac{ Q+(L - a E)^2}{r^3} +\frac{a^2 Q}{r^4}
\end{split}
    \end{equation} where $R(r)$ was introduced in Eq.~\eqref{eq:geodesic-Kerr-Mino}, and we have collected terms in powers of $r^{-1}$. 
As discussed in Sec.~\ref{subsubsec:catastrophe-theory-intro}, the structure of the catastrophe is determined by the order of degeneracy of critical points (given by the conditions on extremizing the generating function in Eq.~\eqref{eq:equilibria-eq-catastrophe}, and its Hessian vanishing~\eqref{eq:hessian}), which translates into the following two conditions for the effective potential
\begin{equation}
    V_{\text{eff}}^\prime =0\,,\quad V_{\text{eff}}^{\prime\prime} =0\,,
\end{equation} with ${}^\prime=\partial_r$. Consequently, the structure of the catastrophe depends critically on the powers of $r^{-1}$ in the effective potential. The highest power in $r^{-1}$ is proportional to $Q$, hence we need to analyze the equatorial ($Q=0$) and non-equatorial ($Q\neq0$) cases separately. 

We start with the equatorial orbits of Kerr, and hence set the Carter constant to zero, for which we obtain a polynomial of degree three in $r^{-1}$. We denote the potential in the equatorial plane by $V_{\text{eff}}^{I=0}$. By using a transformation of M\"{o}bius class~\footnote{A transformation of M\"obius class $r\to \frac{a \mathrm{s}+b}{c \mathrm{s} +d}$ is a local diffeomorphism as long as $ad-bc\neq0$, which is satisfied for the range of parameters under consideration. } 
\begin{equation}\label{eq:Moebiuos-transfo-r}
    r=\frac{6M (L-a E)^2}{a^2 (1-E^2)+L^2-6^{2/3} (L-a E)^{4/3} M^{2/3} \mathrm{s}}
\end{equation} the effective potential for the Kerr radial geodesics in Eq.~\eqref{eq:generating-function-Kerr} confined to the equatorial plane looks like the fold generating function in standard form in Eq.~\eqref{eq:fold-std} up to the constant in Eq.~\eqref{eq:standard form} with
\begin{widetext}
    \begin{equation}
    C_{I=0}(\bm{X}) = \frac{18M^2 (L-a E)^2 [a^2(4E^2-1)-6a E L + 2L^2]+(L^2+a^2(1-E^2))^3}{54 {M^2}(L-a E)^4}
\end{equation}
\end{widetext}
and 
\begin{equation}
    x_1 = \frac{(L^2+a^2(1-E^2))^2-12M^2(L-a E)^2}{ 6^{4/3}(L-a E)^{8/3} M^{4/3}}\,.
\end{equation} Notice that it is not strictly necessary to write the effective potential in the fold standard form to have a fold-like catastrophe manifold. However, we use the standard form of the fold manifold to make the discussion transparent in terms of the general theory of catastrophes introduced in Sec.~\ref{subsubsec:catastrophe-theory-intro}.   

As discussed in Sec.~\ref{subsubsec:catastrophe-theory-intro}, the fold caustic occurs for $x_1=0$, which gives the condition
\begin{equation}\label{eq:condition-caustic}
    \frac{L^2+a^2(1-E^2)}{L-a E} = \pm2\sqrt{3}M\,.
\end{equation} This condition identifies the locus in $(E\,,L)$ space where two stationary points of the effective potential merge. In the Schwarzschild limit this reduces to the usual $L=\pm 2\sqrt{3}M$ condition. 

The two solutions $\mathrm{s} = \pm \sqrt{x_1}$ form the two branches of stationary points of the potential. At the caustic $x_1 = 0$, and so $\mathrm{s}=0$, so that the radius in Eq.~\eqref{eq:Moebiuos-transfo-r} satisfies the relation
\begin{equation}\label{eq:r_l-e}
    r=\sqrt{3}(L-a E)\,,
\end{equation} at the fold caustic (to obtain this result, we used Eq.~\eqref{eq:condition-caustic}). 

Notice that the location of the caustic is still not fully determined in parameter space, since Eq.~\eqref{eq:condition-caustic} provides a relationship between the energy and angular momentum of the orbit for which two stable solutions of the potential coincide. However, to recover the ISCO location we presented in Eq.~\eqref{eq:risco-I=0} we further need to impose the circularity conditions in Eq.~\eqref{eq:circular-condition}, or in this case,  
\begin{equation}\label{eq:circular-condition}
    V_{\text{eff}}^{I=0} =E^2\,.
\end{equation}

Combining Eq.~\eqref{eq:circular-condition} with Eqs.~\eqref{eq:r_l-e} and~\eqref{eq:condition-caustic} we obtain a relationship between the radius and the energy
\begin{equation}
    r=\frac{2M(4\mp3)}{3(1-E^2)}\,,
\end{equation} which can be inverted to obtain an expression of the energy as a function of $r$, and where the $\mp$ stands for prograde and retrograde orbits. Similarly, Eq.~\eqref{eq:r_l-e} can then be inverted. Plugging these expressions in the circular orbit condition in Eq.~\eqref{eq:circular-condition}, we obtain an algebraic equation for the ISCO radial position 
\begin{equation}
r^2 -\left(6M\mp 2a\sqrt{3-\frac{2M}{r}}\right)r+3a^2 =0
\end{equation} whose solutions are given by Eq.~\eqref{eq:risco-I=0}, and where the $\mp$ indicates prograde and retrograde orbits respectively. The above discussion shows that only a given combination of the energy and angular momentum are relevant for the catastrophe morphology. 

Importantly, varying the spin $a$ shifts the location of the fold in parameter space but does not change its codimension or universal structure. The catastrophe type is therefore robust under changes in black hole spin.

Next, we turn to the general case of the space of solutions for inclined orbits in Kerr. The generating function is again the effective potential, given by Eq.~\eqref{eq:def-Veff-explicit}, but since $Q\neq 0$ the polynomial is quartic in $r^{-1}$. As before, we do a local diffeomorphism of the state variables and control parameters to express the Kerr effective potential in the standard form of the catastrophe. In this case, we use the transformation
\begin{equation}\label{eq:diffeo-state-variable-inclination}
    r=\frac{2a^2 Q}{M[(L-a E)^2+Q]+\sqrt{2a^3} Q^{3/4} \mathrm{s}}
\end{equation} which leaves the effective potential in the form
\begin{equation}
    V_{\text{eff}}= C(\bm{X})+\phi_{\text{cusp}}(\mathrm{s}\,;x_1\,,x_2)
\end{equation} where $\phi_{\text{cusp}}$ was given in Eq.~\eqref{eq:cusp-std} and the function $C$ and the control parameters $x_1\,,x_2$ are combinations of $(E\,,L\,,Q)$. The explicit expressions are given in App.~\ref{sec:long-expressions}.  

Given the higher dimensionality of the cusp, we need to study the fold lines and the cusp singularity separately. Here we study the fold lines, since they are the physically relevant quantities for localizing the ISSO in phase space. We briefly comment on the location of the cusp singularity in Sec.~\ref{sec:transition-fold}. 

Following the discussion in Sec.~\ref{subsubsec:catastrophe-theory-intro}, the two fold lines are given by the condition
\begin{equation}
    x_1 = \pm \frac{2}{3} \sqrt{-\frac{x_2^3}{3}}\,,
\end{equation} and correspond to points where 
\begin{equation}
    V_{\text{eff}}^\prime =0\,,\quad V_{\text{eff}}^{\prime\prime} =0\,,\quad V_{\text{eff}}^{\prime\prime\prime} \neq0\,. 
\end{equation}

Solving the Hessian condition in Eq.~\eqref{eq:hessian} yields a relationship between the state variable $\mathrm{s}$ and the control parameter $x_2$,
\begin{equation}\label{eq:fold-lines}
    \mathrm{s} =\pm \sqrt{-\frac{x_2}{3}}\,,
\end{equation} which translates to the radius where the fold singularities would occur through Eq.~\eqref{eq:diffeo-state-variable-inclination}, yielding
\begin{equation} \label{eq:radius-fold}
    r= \frac{2a^2 Q}{M[(L-a E )^2+Q] \pm\sqrt{2a^3}Q^{3/4}\sqrt{-\frac{x_2}{3} }}\,.
\end{equation} 

For quasi-circular orbits, imposing
\begin{equation}\label{eq:quasi-circular}
    V_{\text{eff}}=E^2\,,
\end{equation} restricts the parameter space to a one-dimensional curve that intersects the cusp manifold along its fold lines. Solving the conditions in Eqs.~\eqref{eq:fold-lines},~\eqref{eq:quasi-circular}, and~\eqref{eq:radius-fold} allows to recover the known expressions for the ISSO radius, energy, and angular momentum for fixed spin and inclination that we showed in Figs.~\ref{fig:risco-Kerr-I} and~\ref{fig:Eisco-Kerr-I}.

We have shown that the equilibrium structure of Kerr geodesics is organized by universal catastrophe morphologies: equatorial motion realizes a fold, while generic inclined motion realizes a cusp. The equatorial case is therefore a codimension-one slice of the higher-dimensional cusp manifold. Despite this richer underlying structure, the physically relevant transition to plunge generically corresponds to crossing a fold branch of the cusp rather than its singular tip, which is known as the cusp caustic (see Fig.\ref{fig:cusp-general}). The cusp itself is reached only under non-generic conditions, which we analyze in Sec.~\ref{sec:transition-fold}. As a result, the transition to plunge is structurally stable and universally governed by the slow crossing of the fold, explaining the ubiquity of the Painlevé I differential equation~\cite{Haberman:1979} dominating the dynamics across both equatorial and inclined inspirals.

\subsubsection{The transition as a slow fold crossing}
\label{sec:transition-fold}

In the equatorial case, the transition to plunge corresponds to crossing a fold singularity, and the slow dynamics near this jump is governed by the Painlevé I equation.
However, as reviewed in Sec.~\ref{subseec:transition-to-plunge}, the transition for inclined Kerr orbits is also governed by Painlevé I. This is only consistent if the inspiral trajectory crosses one of the fold branches of the cusp manifold in control parameter space. If instead the cusp singularity itself were encountered, one would expect different universal scaling behavior, associated with the higher catastrophe singularity, which would be governed by the Painlevé II differential equation~\cite{Haberman:1979}.
We now show that crossing the cusp singularity during the transition occurs only under non-generic conditions.

The cusp singularity occurs where both fold lines coincide, i.e., 
\begin{equation}\label{eq:caustic-control-params}
    x_1=x_2 =0\quad \Rightarrow \quad\mathrm{s} =0\,.
\end{equation}

Using the diffeomorphism introduced in Eq.~\eqref{eq:diffeo-state-variable-inclination}, this corresponds to
\begin{equation}
r=\frac{2a^2 Q}{M[(L-a E)^2+Q]} .
\end{equation}

To determine whether this configuration is realized dynamically, it is convenient to return to the radial polynomial $R(r)$ in its factorized form in Eq.~\eqref{eq:R(r)-bounded}. For quasi-circular orbits,
\begin{equation}
\partial_r V_{\text{eff}}=0 \quad \Rightarrow \quad \partial_r R=0,
\quad
\partial_r^2 V_{\text{eff}}=0 \quad \Rightarrow \quad \partial_r^2 R=0 ,
\end{equation}
so the critical points coincide for both functions, and we can use $R$ for the following analysis.

The ISSO corresponds to a triple root of the function $R$,
\begin{equation}
R(r_{\text{ISSO}})=R'(r_{\text{ISSO}})=R''(r_{\text{ISSO}})=0 ,
\end{equation}
i.e. $r_1=r_2=r_3=r_{\text{ISSO}}$. This is one of the fold lines of the cusp manifold. The second fold line is not crossed on the transition setting, since it would correspond to having a triple root inside the horizon $r_1 = r_2=r_4\equiv r_{\text{fold}}\neq r_3$ with $r_4 = \frac{a^2 Q}{(1-E^2)r_3 r_{\text{fold}}}$. However, a cusp singularity would require an additional degeneracy, namely
\begin{equation}
r_4 = r_{\text{ISSO}}\equiv r_{\text{cusp}} \,,
\end{equation}
so that $R(r)$ develops a root of multiplicity four
\begin{equation}
R(r)=-(1-E^2)(r_{\text{cusp}}-r)^4 \,.
\end{equation}

Substituting this ansatz into Eq.~\eqref{eq:def-Veff-explicit} and matching powers of $r$ yields an algebraic relation between the dimensionless spin $\chi=a/M$ and $q=Q/M^2$:
\begin{widetext}
\begin{equation}\label{eq:condition-cusp}
    \chi^2 + \chi\sqrt{4-\frac{2^{2/3}}{\chi^{2/3} q^{1/3}}}\sqrt{2^{2/3} \chi^{4/3} q^{2/3}-q}-3 2^{1/3} \chi^{2/3} q^{1/3} +2^{2/3} \chi^{4/3} q^{2/3} = 0\,. 
\end{equation}    
\end{widetext}
 Although this equation would in principle generate a family of black hole spins and orbit inclinations for which the cusp crossing could be possible, one can show (see App.~\ref{sec:proof-subextremal-cusp}) that this equation admits no real solution for $0<\chi<1$. A real solution exists only in the extremal limit $\chi=1$, for which $Q/M^2=1/2$~\footnote{This value of the Carter constant corresponds to $I\approx24^\circ$ for prograde orbits in the extremal Kerr limit. This inclination angle coincides with $\theta_{\text{inc, kink}}^{\text{sph.}}$ in Ref.~\cite{Stein_2020}.}.

Thus, the cusp singularity arises only in the extremal limit and for a specific value of the inclination. In other words, in control-parameter space it lies on a codimension-two surface, and is therefore non-generic: reaching it requires fine-tuning of two independent conditions.  

A more detailed analysis of the near-horizon geometry in the extremal Kerr limit{, analogous to the one in Ref.~\cite{Compère_2020-Painleve},} would be required to determine whether, in that regime, the transition to plunge is governed by a different universal scaling (in particular, whether it could be described by the Painlevé II equation), as suggested by the higher degeneracy of the potential. Establishing this would also clarify whether the cusp can indeed be accessed only for a finely tuned inclination.

We therefore interpret the above discussion in two ways. First, it limits the validity of our derivation to subextremal black holes. Second, it reinforces the universality of the transition being governed by the Painlevé I equation for arbitrary inclination and spin, provided that $\chi<1$.

If, however, the transition to plunge in an extremal Kerr background were controlled by a different dynamical scaling, this would constitute a qualitatively distinct observational signature. Such behavior could, in principle, provide a means of distinguishing extremal black holes, should they exist in nature.

The discussion above shows that a generic transition to plunge does not cross the cusp singularity. We must still justify two points: (i) radiation reaction restricts the evolution to an effectively one-dimensional slice of the cusp manifold, and (ii) the trajectory crosses the fold branch transversely, so the local dynamics follows the fold normal form and is therefore governed by the Painlevé I equation.

Near the fold lines, radiation reaction drives the system along a one-dimensional trajectory in control-parameter space. This follows from the fact that, during the transition regime discussed in Sec.~\ref{subseec:transition-to-plunge}, the quantities $(E,L,Q)$ share the same scaling with $(\lambda-\lambda_{\text{ISCO}})$. The inspiral therefore traces a one-dimensional curve in the two-dimensional control space $(x_1\,,x_2)$.

The only remaining possibility to exclude is that this curve lies tangent to the fold branch, rather than crossing it. Away from the cusp tip, however, any one-dimensional slicing of the cusp manifold is locally diffeomorphic to a fold. Thus, the precise ``angle'' of crossing is irrelevant, what matters is that the trajectory does not remain confined to the fold branch.

Transversality can be expressed by requiring that the total derivative of the fold constraint along the inspiral trajectory be nonvanishing at the crossing point. If the fold branches are given by Eq.~\eqref{eq:fold-lines}, then transversality requires
\begin{equation}\label{eq:transversality}
    \frac{\d}{\d\lambda}\left(x_1\mp \frac{2}{3} \sqrt{-\frac{x_2^3}{3}}\right)\Bigg|_{x_1\pm \frac{2}{3} \sqrt{-\frac{x_2^3}{3}}=0} \neq 0 \, .
\end{equation} 
Equivalently, the velocity vector in control space must not be tangent to the fold curve.

Since radiation reaction generically induces independent variations of the control parameters, the resulting evolution in control-parameter space is not expected to remain tangent to the fold line. Such tangential motion would require a fine tuning of the fluxes (equivalently, of the coefficients $\kappa_{\mathcal{C}}$ in Eqs.~\eqref{eq:delta-c} and~\eqref{eq:C-evolution}) and is therefore non-generic. Furthermore, we have explicitly checked that Eq.~\eqref{eq:transversality} is satisfied for the values given in Tab.~\ref{tab:kappas-A-B}. 

Therefore, although the full equilibrium manifold for inclined Kerr orbits has cusp morphology, the inspiral trajectory generically crosses one of its fold branches transversely. By structural stability, the local dynamics is then governed by the universal fold normal form.

This explains why the transition to plunge for inclined Kerr orbits is described by the Painlevé I equation despite the underlying cusp structure of the equilibrium manifold.

\section{Discussion}
\label{sec:discussion}

Accurately modeling the transition from adiabatic inspiral to plunge is essential for constructing faithful IMRI/EMRI waveform models. A central result of this work is that this regime exhibits a striking universality: for quasi-circular orbits in rotating black-hole spacetimes, both equatorial and inclined, the transition dynamics is governed by the Painlev\'e~I equation. Our aim has been to elucidate the origin of this universality in the presence of the additional structure introduced by inclination. Within a catastrophe-theory framework we found that, while inclined orbits enrich the geometry of the equilibrium manifold (as already suggested by the greater complexity of the ISSO relative to the ISCO), the physical inspiral nevertheless selects a fold branch, effectively reducing the dynamics to the equatorial case. Finally, we showed that the Painlev\'e~I description is structurally stable, in the sense that it persists under perturbations.

{ Extending the work in \cite{Jaramillo:2022oqn}}, we also showed that
the \emph{tritronquée solution}, defined by its asymptotic behavior in
the complex plane, is the relevant physical solution to the Painlev\'e I equation. We
demonstrated that the known high-accuracy analytical solution is at least as accurate as current numerical methods. Moreover, there exist uniform error bounds for this solution, making it practically useful for quantitative studies of this ``merger'' regime. If necessary, the analytic solution could be made more accurate by extending the methods of Ref.~\cite{ADALITANVEER20163843}.

The application of catastrophe theory in the context of binary mergers was first put forward for comparable-mass systems in \cite{Jaramillo:2022mkh}, and is investigated here in greater detail in the EMRI setting. 
{ The present study centered on the inspiral-to-plunge transition, with its use of approximate analytic solutions and its quest for structurally stable structures, provides a remarkable instance of the ``asymptotic reasoning'' \cite{Batte01} approach  in \cite{Jaramillo:2022mkh,Jaramillo:2023day,Jaramillo:2022oqn,JarAlSBes26}}. { In particular, the present work provides a non-trivial case of insight gained into the structural mechanisms that underlie the simplicity and universality of binary black hole merger dynamics.
}
%, namely aiming at gaining insights into the mechanisms that underlie the simplicity and universality of binary black hole merger dynamics}.

Natural, though non-trivial, extensions of this work include generalizing the analysis to eccentric equatorial inspirals and, ultimately, to fully generic (inclined and eccentric) Kerr orbits. It would also be interesting to incorporate higher-order effects (such as higher order self-force and post-leading transition-to-plunge orders), 
and to translate the resulting dynamics into practical waveform ingredients (e.g. for fast surrogate models, reduced-order models or Fast EMRI Waveforms (FEW) \cite{Chua:2020stf,Katz:2021yft}). 
Finally, one could explore how robust the catastrophe-theory picture remains in the presence of additional physics, such as small-body spin, environmental perturbations, or beyond-GR modifications that deform the separatrix structure.

\section*{Acknowledgments}

ARM would like to thank Maarten van de Meent for helpful discussions on this project and for providing the Mathematica interpolants used in the construction of Tab.~\ref{tab:kappas-A-B}. ARM and BB thank Leor Barack for stimulating discussions and all authors thank Leor Barack, Lorenzo Küchler, and Geoffrey Comp\`ere for feedback on a draft version of this paper.
ARM is also grateful to Conor Dyson, and Lorenzo Küchler for insightful discussions and for sharing the parameter choices used in the numerical integrations performed in Mathematica. 
ARM further thanks Scott Hughes and Devin Becker for providing their transition-to-plunge data, which enabled consistency checks of the exact solution in the early stages of the project.
ARM and BB thank Patrick Bourg for providing data for the rate of change of the constants of motion using FEW.  
{ JLJ would like to thank Ricardo Uribe-Vargas and Oscas Meneses-Rojas for the many discussions and their insights in catastrophe (singularity) theory.}
This work makes use of the Black Hole Perturbation Toolkit \cite{BHPToolkit}.  The Center of Gravity is a Center of Excellence funded by the Danish National Research Foundation under grant No. 184.
 { PO FEDER-FSE Bourgogne 2014/2020 program and the EIPHI Graduate School (contract ANR-17-EURE- 0002) as part of the ISA 2019 project}. 
Funded by the European Union (ERC grant GWSky/ 101167314). Views and opinions expressed are however those of the author(s) only and do not necessarily reflect those of the European Union or the European Research Council Executive Agency. Neither the European Union nor the granting authority can be held responsible for them.

\bibliographystyle{apsrev4-1}
\bibliography{main}
\appendix

\section{Coefficients and special functions for the approximate tritronqu\'ee solution }
\label{sec:extra-approximate-sol}
The tritronqu\'ee solution of the Painlev\'e~I (PI) differential equation in Eq.~\eqref{eq:PainleveI}, with $T\in\mathbb{C}$, is the unique solution satisfying the asymptotic behavior
\begin{equation}
    X(T)=\sqrt{-T}\left[1+\mathrm{O}(T^{-5/8})\right]
    \qquad \text{as } T\to -\infty\,.
\end{equation}
Any solution of PI is single-valued and meromorphic in $\mathbb{C}$, with the singularity locations determined by the initial data $(X,\dot{X})$.

Rather than specifying initial conditions, a solution can equivalently be characterized by the location of a pole $T_{\mathrm{pole}}$ together with the coefficient $\widehat{a}_2$ appearing in its local series expansion. The solution then admits the locally convergent representation
\begin{widetext}
\begin{equation}\label{eq:exact-series-expansion}
    X(T)
    =
    -\frac{6}{(T-T_{\mathrm{pole}})^2}
    -6^{2/5}(T-T_{\mathrm{pole}})^2
    \sum_{j=0}^{\infty}
    (-1)^j
    \frac{\widehat{a}_j}{6^{j/5}}
    (T-T_{\mathrm{pole}})^j \,.
\end{equation}
\end{widetext}
The coefficients satisfy
\begin{equation}
    \widehat{a}_0=\frac{T_{\mathrm{pole}}}{6^{1/5}10}\,,
    \qquad
    \widehat{a}_1=-\frac{1}{6}\,,
    \qquad
    \widehat{a}_3=0\,,
\end{equation}
while for $n\geq4$ they are determined recursively through
\begin{equation}\label{eq:exact-series-ak}
    \widehat{a}_n
    =
    -\frac{6}{(n+5)(n-2)}
    \sum_{k=0}^{n-4}
    \widehat{a}_k
    \widehat{a}_{n-4-k}\,.
\end{equation}

The exact location of the first singularity on the positive real axis, $T_{\mathrm{pole}}$, is not known analytically. However, the approximation $T_0$ given in Eq.~\eqref{eq:T0} differs from the true pole location by at most $\mathrm{O}(10^{-6})$.

In Sec.~\ref{sec:tritronque-sol} we introduced the approximate solution constructed in Ref.~\cite{ADALITANVEER20163843}, which replaces the exact pole location $T_{\mathrm{pole}}$ by the approximation $T_0$ while still providing rigorous error bounds for both the solution and its derivative. In the remainder of this appendix we summarize the coefficients and auxiliary functions entering the approximate solution $X_0$ defined in Eq.~\eqref{eq:approx-exact-sol-X0}.

The auxiliary functions $P_2$ and $P_3$ introduced in Eq.~\eqref{eq:P2-P3} are finite-order polynomials. The coefficients $b_k$ are given in Eq.~(7) of Ref.~\cite{ADALITANVEER20163843}. The coefficients $c_k$ are defined analogously to those in Eqs.~\eqref{eq:exact-series-expansion} and~\eqref{eq:exact-series-ak}, but replacing $T_{\mathrm{pole}}$ by $T_0$ and fixing the coefficient $c_2$ to
\begin{equation}
    c_2=\frac{19949}{321055}\,.
\end{equation}
The remaining coefficients $c_n$, with $n=4,\ldots,17$, are then determined recursively through Eq.~\eqref{eq:exact-series-ak}.

Finally, the function $\mathcal{W}_0$ introduced in Eq.~\eqref{eq:wo} is given explicitly by
\begin{widetext}
\begin{equation}
\mathcal{W}_{0}(z,s)
=
-\frac{4412401}{245760 z^7 (54)^{1/4}}
\left[
(1+i s)^{-15/2}
-\frac{1225\sqrt{6}}{540294 z^2}(1+i s)^{-19/2}
+\frac{30625}{2161176 z^4}(1+i s)^{-23/2}
\right].
\end{equation}
\end{widetext}

\label{sec:Painleve-trascendents}

\section{Long expressions omitted in Sec.~\ref{subsec:catastrophe-theory}}
\label{sec:long-expressions}

In Sec.~\ref{subsec:Kerr-parameter-space}, we omitted the explicit expressions for the control parameters $(x_1,x_2)$ and the function $C(\mathbf{X})$ appearing in the reduction of the Kerr effective potential for inclined orbits to the cusp normal form~\eqref{eq:cusp-std}, since these expressions are rather lengthy. For completeness, we provide them here.

The control parameters entering the cusp generating function are
\begin{widetext}
    \begin{subequations}
        \begin{equation}
            x_1= -\frac{M}{2 \sqrt{2} a^{9/2} Q^{9/4}}  \left[4 a^4 Q^2 + 
   2 a^2 (a^2 (-1 + E^2) - L^2 - Q) Q ((-a E + L)^2 + Q) + 
   2 M^2 ((-a E + L)^2 + Q)^3)\right]\,,
        \end{equation}
        \begin{equation}
            x_2 = \frac{1}{2 a^3 Q^{
  3/2}} \left(-3 (-aE + L)^4 M^2 + 
   2 (-a^4 (-1 +E^2) + a^2 L^2 - 3 (-aE + L)^2 M^2) Q + (2 a^2 - 
      3 M^2) Q^2\right)\,.
        \end{equation}
       The additional smooth contribution appearing in Eq.~\eqref{eq:standard form} is
        \begin{equation}
            C(\bm{X}) = 1 - \frac{M^2 ((-a E + L)^2 + Q)}{a^2 Q} - \frac{
 M^2 (a^2 (-1 + E^2) - L^2 - Q) ((-a E + L)^2 + Q)^2}{4 a^4 Q^2} - \frac{
 3 M^4 ((-a E + L)^2 + Q)^4}{16 a^6 Q^3}\,.
        \end{equation}
    \end{subequations}
\end{widetext}

\section{Proof that the cusp singularity is not crossed for subextremal black holes}
\label{sec:proof-subextremal-cusp}

The purpose of this appendix is to determine whether Eq.~\eqref{eq:condition-cusp} admits real solutions. It is immediate to verify that $\chi=0$ formally solves the equation. However, the non-rotating case was treated separately in Sec.~\ref{sec:transition-fold}, where the transition is governed by a fold rather than a cusp. Eq.~\eqref{eq:condition-cusp} therefore does not apply to the non-rotating case , so $\chi>0$ is strictly larger than zero. For $\chi=1$, Eq.~\eqref{eq:condition-cusp} admits a unique solution corresponding to $q=1/2$, as discussed in Sec.~\ref{sec:transition-fold}. We now show that no additional real solutions exist for $0<\chi<1$.

To analyze the equation it is convenient to introduce the change of variables
\begin{equation}
x=2^{1/3}\chi^{2/3} q^{1/3},
\end{equation}
which eliminates $q$. In terms of $\chi$ and $x$, Eq.~\eqref{eq:condition-cusp} becomes
\begin{equation}\label{eq:chi-equation}
F(\chi\,,x) \equiv \chi^2
+ x\chi
\sqrt{2-\frac{1}{x}}
\sqrt{2-\frac{x}{\chi^2}}
-3 x
+x^2
=0.
\end{equation}

For real solutions, the arguments of both square roots must have the same sign, so either both are positive or both are negative.  Hence
\begin{equation}
\left(2-\frac{1}{x}\right)
\left(2-\frac{x}{\chi^2}\right)\ge0.
\end{equation}
The branch in which both factors are negative is incompatible with $0<\chi\le1$, so we require both to be positive. This yields the domain restriction
\begin{equation} 
\frac{1}{2}
\le x
\le
2\chi^2.
\end{equation}
For this interval to be non-empty, the upper bound must exceed the lower bound, which requires
\begin{equation}
2\chi^2
\ge
\frac{1}{2}
\quad\Longrightarrow\quad
\chi\ge\frac{1}{2}.
\end{equation}
Therefore no real solutions exist for slowly spinning black holes with
$\chi<1/{2}$.

It remains to analyze the regime $1/{2}\le\chi<1$. Since a solution exists at extremality, we treat the near-extremal case perturbatively by setting
\begin{equation}
\chi=1-\epsilon,
\qquad
\epsilon\ll1.
\end{equation}
At $\chi=1$, the solution lies at the lower endpoint of the allowed interval, $x=1$. We therefore consider the perturbative expansion
\begin{equation}
x=1
-x_1\,\epsilon,
\end{equation}
with $x_1$ an $\mathrm{O}(1)$ parameter. Substituting into Eq.~\eqref{eq:chi-equation} and expanding to linear order in $\epsilon$ yields
\begin{equation}
-4\epsilon+\mathrm{O}(\epsilon^2).
\end{equation}
This expression is independent of $x_1$ and vanishes only in the limit $\epsilon\to0$. Consequently, no nearby real solutions bifurcate away from the extremal point.

We are just left with arguing that no real solutions exist for $1>\chi>\frac{1}{{2}}$ and $x\in(1/2\,,2\chi^2)$. To do this, we show that the function $F(\chi\,,x)$ defined in Eq.~\eqref{eq:chi-equation} is monotonic and that it remains negative in the above interval.

For $x=1$, $F(\chi\,,x)$ simplifies to 
\begin{equation}
    F(\chi\,,{1}) = -2+\chi\sqrt{2-\frac{1}{\chi^2}} +\chi^2
\end{equation} which remains negative and has the limit
\begin{equation}
    \lim_{\chi\to1^-} F(\chi\,,{1}) \to0\,.
\end{equation} For the upper boundary $x=2\chi^2$ the function simplifies to 
\begin{equation}
    F(\chi\,,{2\chi^2}) = \chi^2 (-5 + 4 \chi^4)\,,
\end{equation} but since $\chi$ is strictly smaller than one, the function remains negative for all allowed $\chi$ on the interval. Similarly, one can show that the function $F(\chi\,,x)$ contains no stationary points in the above interval, so $F(\chi\,,x)$ is monotonic and remains negative in the interval $1>\chi>\frac{1}{{2}}$ and $x\in(1/2\,,2\chi^2)$. Hence, we conclude that Eq.~\eqref{eq:condition-cusp} admits a real solution only in the extremal case $\chi=1$, and no additional real solutions exist for $0<\chi<1$.
\end{document}